\documentclass{aa}  

\usepackage{graphicx}
\usepackage{lscape}
\usepackage[varg]{txfonts}

\usepackage{color}
\newcommand{\gsim}{\hbox{\rlap{\lower.55ex\hbox{$\sim$}} \kern-.3em
\raise.4ex \hbox{$>$}}}
\newcommand{\lsim}{\hbox{\rlap{\lower.55ex\hbox{$\sim$}} \kern-.3em
\raise.4ex \hbox{$<$}}}

\newcommand{\ha}{H$\alpha$}

\newcommand{\hii}{\textsc{H\,ii}}
\newcommand{\hei}{He\,\textsc{i}}

\newcommand{\oi}{[O\,\textsc{i}]}

\newcommand{\ov}{O\,\textsc{v}}

\newcommand{\feiii}{[Fe\,\textsc{iii}]}

\newcommand{\ciii}{C\,\textsc{iii}}
\newcommand{\civ}{C\,\textsc{iv}}
\newcommand{\mgi}{Mg\,\textsc{i}]}
\newcommand{\ariv}{[Ar\,\textsc{iv}]}

\newcommand{\nii}{\textsc{[N\,ii]}}

\newcommand{\niii}{\textsc{N\,iii}}
\newcommand{\nv}{\textsc{N\,v}}

\newcommand{\heii}{He\,\textsc{ii}}

\def\pthreed{\textsc{p3d}}

\begin{document}
%
   \title{The Wolf-Rayet star population in the dwarf galaxy \object{NGC 625}}


   \author{A. Monreal-Ibero
          \inst{1,2,3,4}
          \and
          J. R. Walsh
          \inst{5}
          \and
          J. Iglesias-P\'aramo
          \inst{4,6}
          \and
          C. Sandin
          \inst{3}
          \and
          \\
          M. Rela\~no
          \inst{7,8}
          \and
          E. P\'erez-Montero
          \inst{4}
          \and
          J. V\'{\i}lchez
          \inst{4}
          \fnmsep\thanks{Based on observations
       collected at the European Organisation for Astronomical
       Research in the Southern Hemisphere, Chile (ESO Programme
       086.B-0042).}
          }
   \institute{
Instituto de Astrof\'{\i}sica de Canarias (IAC), E-38205 La Laguna, Tenerife, Spain
\and
Universidad de La Laguna, Dpto. Astrof\'{\i}sica, E-38206 La Laguna, Tenerife, Spain
\email{amonreal@iac.es}
\and
Leibniz-Institut f\"ur Astrophysik Potsdam (AIP), An der Sternwarte 16, 14482 Potsdam, Germany
\and
Instituto de Astrof\'{\i}sica de Andaluc\'{\i}a (CSIC), C/  Camino Bajo de Hu\'etor, 50, 18008 Granada, Spain
\and 
European Southern Observatory. Karl-Schwarzschild Strasse 2, 85748 Garching, Germany 
\and
Estaci\'{o}n Experimental de Zonas \'{A}ridas (CSIC), Ctra. de Sacramento s/n, La Ca\~{n}ada, Almer\'{\i}a, Spain
\and
Dept. F\'{\i}sica Te\'orica y del Cosmos, Universidad de Granada, 18010 Granada, Spain 
\and
Instituto Universitario Carlos I de F\'{\i}sica Te\'orica y Computacional, Universidad de Granada, 18071 Granada, Spain      
             }

\date{Received DD/MM/YYYY; accepted DD/MM/YYYY}

 
  \abstract
   {Quantifying the number, type and distribution of Wolf-Rayet (W-R) stars is a key component in the context of galaxy evolution, since they put constraints on the age of the star formation bursts. Nearby galaxies (distances $\lsim$5~Mpc) are particularly relevant in this context since they fill the gap between studies in the Local Group, where individual stars can be resolved, and galaxies in the Local Volume and beyond.}
   {We intend to characterize the W-R star population in one of these
systems: \object{NGC~625}, a low-metallicity dwarf galaxy suffering a currently declining burst of star formation.}
   {Optical integral field spectroscopy (IFS) data have been obtained with the
VIMOS-IFU and the HR\_Orange and HR\_Blue gratings at the Very Large Telescope covering the starburst region of \object{NGC\,625}.
Ancillary HST images in the $F555W$ and $F814W$ bands are also used for comparison.
We estimate the number of W-R stars using a linear combination of three W-R templates: one early-type nitrogen (WN) star,  one late-type WN star and one carbon-type (WC)  star (or oxygen-type (WO) star). Fits using several ensembles of templates were tested. Results were confronted with: i) high spatial resolution HST photometry; ii) numbers of W-R stars in nearby galaxies; iii) model predictions.
}
   {
The W-R star population is spread over the main body of the galaxy, not necessarily coincident with the overall stellar distribution.
Our best estimation for the number of W-R stars yields a total of 28 W-R stars in the galaxy, out of which 17 are early-type WN, six are late-type WN and five are WC stars, with our most reasonable estimation.
The width of the stellar features nicely correlates with the dominant W-R type found in each aperture.
The distribution of the different types of WR in the galaxy is roughly compatible with the way star formation has propagated in the galaxy, according to previous findings using high spatial resolution with the HST.
Fits using templates at the metallicity of the Large Magellanic Cloud yield more reasonable number of W-R than those using templates at the metallicity of the Small Magellanic Cloud. Given the metallicity of \object{NGC\,625}, this suggests a non-linear relation between the metallicity and the luminosity of the W-R spectral features.
}
   {}

 
   \maketitle
%

\section{Introduction}

Wolf-Rayet (W-R) stars constitute a stage in the evolution of
massive (M$>$25~M$_\odot$) stars once they leave the main sequence \citep[see ][for a
review]{cro07}. They are understood as stars that have lost the bulk of their
hydrogen-rich envelope through stellar winds and are currently burning helium in
their core \citep[e.g.][]{con75}.
They are luminous objects with strong broad emission lines in their spectra.
In a very simplified manner, they 
can be classified as nitrogen  (WN) stars (those with strong lines of helium and
nitrogen) and carbon (WC and WO)
stars (those with strong lines of helium, and carbon or oxygen). They
lose a significant amount of their mass via their dense stellar winds and show
the products of the CNO-burning first (WN stars) and the He-burning afterwards
(WC and WO stars).

There are several reason why these stars are particularly relevant in the
context of galaxy evolution. 
Firstly, they can contribute significantly to the chemical enrichment of
galaxies \citep[e.g.][]{dra03,mae92}.
Secondly, given the relatively short duration of the W-R phase, the detection of
spectral features typical of these stars constitutes a very precise method for
estimating the age of a given stellar population.
For example, instantaneous bursts of star formation show these features at ages of $\sim2-6$~Myr  \citep{Leitherer99}. 
Also, they are candidate progenitors for Type Ib/Ic supernovae \citep{sma09} and
possible progenitors of long, soft gamma-ray bursts \citep{woo06}.

Wolf-Rayet galaxies are those with a spectrum
displaying features typical of Wolf-Rayet stars \citep{con91,sch99}. 
Because of this definition, the classification of an object as W-R galaxy may depend on the way the object was observed and thus is not a well defined category. Most of the times it refers to a galaxy where the spectra of (some of) its \hii\, regions display W-R features but also, it might refers to galaxies that display these features in their integrated spectrum.
The most important W-R features in these galaxies  are the bumps around 4650\AA\
(i.e. the \emph{blue bump}, mainly but not always characteristic of WN stars) and 5808 \AA\ (i.e. the \emph{red
bump}, characteristic of WC stars). Both of them are blends made out of several narrow
nebular and broader stellar emission lines.
Without constituting an homogeneous class, W-R galaxies share (at least) one common characteristic:
 because of the nature of W-R stars, they have ongoing or recent star formation which has produced
stars massive enough to evolve to the W-R stage. This indicates stellar
populations with typical ages of $\lsim$10~Myr.
The first galaxy where W-R emission was detected was He 2-10, a Blue Compact
Dwarf (BCD) galaxy \citep{all76} but W-R galaxies can be found among a large
variety of galaxy types. This includes not only low-mass BCD and
irregular galaxies \citep[e.g.][]{izo97,lop10}, but also massive spirals
\citep[e.g.][]{phi92,fer04} and luminous infrared galaxies
\citep[e.g.][]{arm88,lip03}.

The study of the W-R star
population in very nearby (i.e. \lsim5~Mpc) starbursts and regions of star
formation \citep[e.g.][]{Hadfield07,dri99,cro09,bib10} is specially relevant since it acts
as a  bridge  filling the gap between the study of individual W-R stars and W-R
galaxies at larger distances, where the locations of recent star formation cannot be spatially resolved \citep[e.g.][]{bri08}.

\begin{figure}[th!]
   \centering
\includegraphics[angle=0,width=0.48\textwidth, clip=,bb = 45 30 480 255]{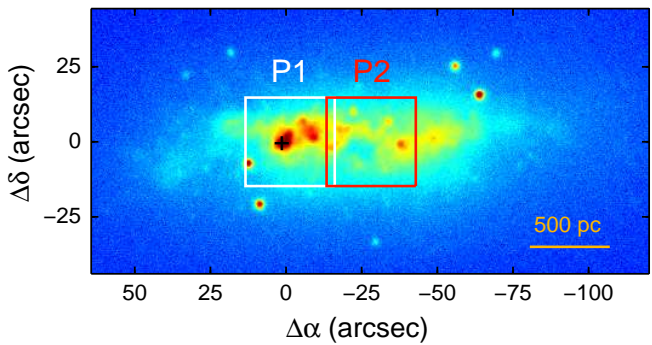}
   \caption[Covered field with VIMOS]{$B$ band image of NGC~625 from the Ohio
State University (OSU) Bright Spiral Galaxy Survey \citep{esk02}. The area
covered in ESO program
     086.B-0042 is marked with two squares (white for pointing 1, red for pointing 2).
     The scale for a distance of 3.9~Mpc is indicated in the lower right corner. The orientation is north (up) and east (to the left). 
 \label{apuntado2}}
 \end{figure}

\object{NGC~625} constitutes an example in this category,
 since W-R features have been detected in its main \hii\ region
\citep{ski03}. This irregular galaxy with moderately low metallicity (see Table \ref{datosbasicos}) is a member of the Sculptor Group, the
nearest group of galaxies outside the Local Group \citep{kar03}. 
Also, according to the criteria proposed by \citet{gil03} and \citet{thu81}, it can
be classified as a BCD.
The star-formation history  of the galaxy is well characterised via
detailed photometry of individual stars using WFPC2 images \citep{can03,mcq10b}.
The galaxy has a well-defined radial stellar population gradient. It suffers a
(currently declining) burst of SF, lasting
$\sim$450~Myr, with the most elevated rates of star formation at $<100$~Myr.

We are currently doing a detailed study of this galaxy using deep Integral Field
Spectroscopy observations of its $\sim$1\,100~pc$\times$550~pc central area with the VIMOS-IFU. 
As first illustrated by \citet{bas06}, Integral Field Spectroscopy provides 
a swift way for the detection and characterization of W-R emission. Using the same set
of observations, one can identify the locations with W-R emission by simulating the action of narrow filters and
creating continuum-subtracted maps at the \emph{blue} and/or \emph{red} bumps, and localizing the peaks of emission
afterwards. Then, the emission can be characterised by extracting the spectra of the spaxels associated to the peaks.
This technique has been used in a routine manner in recent years.
For example, recently the first catalogue of W-R star rich regions with spatially-resolved information has been released to the community in the context of the CALIFA survey  \citep{MirallesCaballero16} 
%
%
and the pool of examples of nearby (i.e. $\lsim5$~Mpc)
galaxies whose W-R population is being
detected and/or characterised by means of this methodology is rapidly increasing
\citep[e.g.][]{rel10,MonrealIbero10a,lop10,mon11,Westmoquette13,keh13}.
In this paper, we will present the VIMOS-IFU observations and the characterization of the
Wolf-Rayet star population in \object{NGC~625}.

The paper is organized as follows: Sect. \ref{secobservations} describes the
characteristics of the observations; Sect. \ref{secreduction} contains the
technical details regarding the data reduction and derivation of the required
observables; Sect. \ref{secresults} characterise the W-R star population in the galaxy; Sect. \ref{secdiscu} compares our results with
previous findings using high spatial resolution with the HST, with the W-R content in galaxies at similar distances and with the predictions of stellar models.
Our main conclusions and perspectives for future investigations are summarised in Sect. \ref{secconclu}.
Basic information for \object{NGC~625} can be found in Table~\ref{datosbasicos}. 



 


%
\begin{table}
\caption{Basic data for \object{NGC~625}}             
\label{datosbasicos}      
\centering                      
\begin{tabular}{c c c c}        
\hline\hline                    
Parameter & Value  & Ref.\\           
\hline                          
Name               & NGC~625 & (a)\\ 
Other  		   & ESO 297- G 005,   & (a)\\
Designations       & IRAS F01329-4141, & \\
                   & AM 0132-414    & \\
Hubble type        & SB(s)m sp & (b) \\
RA (J2000.0)       & 01h35m04.6s  & (a)\\
Dec (J2000.0)      & -41d26m10s   & (a)\\
z                  & 0.001321     & (a)\\
$D({\rm Mpc})$ 	   & 3.9$\pm$0.2 & (c)\\
Scale (pc/$^{\prime\prime}$) &  18.9	&\\
$12+\log(O/H)$     & 8.14$\pm$0.02 & (d)\\
$E(B-V)_{gal}$     & 0.015       & (e) \\
$M_{\rm HI} (M_\odot)$ & $1.1\times10^8$ & (f)\\
$M_B$              & $-16.20$ &  (g,$\ast$)\\
$<\mu_{B,core}>$ (mag/arcsec$^{2}$)   & 21.15  & (g)\\
$(B-V)_{core}$     & 0.22     &  (g)\\ 
$(V-I)_{core}$     & 0.49     &  (g)\\ 
$\log(L_{\rm fir}/L_\odot)$ 	& 8.28	&(h,$\ast$)\\ 
$\log(L_{\rm ir}/L_\odot)$ 	& 8.45	&(h,$\ast$)\\ 
$m_{FUV}$              & 13.72$\pm$0.05 & (i)\\
$m_{NUV}$              & 13.34$\pm$0.03 & (i)\\
$m_H$              & 8.94$\pm$0.04  & (j)\\
$M_H$              & $-19.01\pm0.2$  & (j,$\ast$)\\
$H-K$              & 0.19           & (k) \\ 
$\log(M_\ast/M_\odot)$ &  8.3$\pm$0.2 & (j)\\
\hline                                   
\end{tabular}
\begin{flushleft}
References:
$^{{\rm (a)}}$ NASA/IPAC Extragalactic Database (NED);
$^{{\rm (b)}}$ Third Reference Catalog of Bright Galaxies \citep[RC3][]{dev91};
$^{{\rm (c)}}$ \citet{can03};
$^{{\rm (d)}}$ \citet{ski03};
$^{{\rm (e)}}$ \citet{sch11};
$^{{\rm (f)}}$ \citet{can04b};
$^{{\rm (g)}}$ \citet{mar97};
$^{{\rm (h)}}$ \citet{san03};
$^{{\rm (i)}}$ \citet{lee11};
$^{{\rm (j)}}$ \citet{kir08};
$^{{\rm (k)}}$ \citet{jar03};
$^{(\ast)}$ Re-scaled to the distance adopted here;
\end{flushleft}
\end{table}

\section{Observations \label{secobservations}}

Data were obtained in Service Mode during Period 86 using the integral field
unit of VIMOS \citep{lef03} at the Unit 3 (Melipal) of the Very Large Telescope
(VLT). Both the HR\_Blue and HR\_Orange grisms were used. Their nominal spectral
ranges are $4\,150 - 6\,200$~\AA\ and $5\,250 - 7\,400$~\AA, respectively,
allowing to 
cover in a continuous manner from 4\,200~\AA\ to 7\,200~\AA. In these modes the
field of view per pointing covers $27^{\prime\prime}\times27^{\prime\prime}$
with a magnification set to 0\farcs67 per spatial element (hereafter,
\emph{spaxel}). 

In order to fully map the area of strongest level of star formation a mosaic of
two tiles was needed with an offset of 29\farcs0 (i.e. 44 spaxels) in right
ascension between them (see Fig. \ref{apuntado2}). For each tile, a square 4 pointing dither pattern with a
relative off-set of 2\farcs7, equivalent to 4 spaxels, was used. It has been
proved that this pattern is adequate to minimize the effect of dead fibers  and
therefore map continuously the area of interest \citep[][]{arr08}. Aditionally
after finishing the observation of each square pattern, a nearby sky background
frame was obtained to evaluate and subtract the background emission.
 Standard sets of calibration files were also obtained. These included a set of 
  continuum and arc lamps exposures.
Each set of dither pattern,
background and standard calibration exposures constituted an observing block.
Finally, spectrophotometric standards were also observed as part of the calibration plan of the observatory.

Atmospheric conditions during the observations were clear and the typical seeing
ranged between 0\farcs4 and 1\farcs1. All the data were taken at low airmasses
in order to prevent strong effects due to differential atmospheric refraction.
Nevertheless, as a safety test, the magnitude of these effects was checked a
posteriori during the reduction process (see Sect. \ref{secreduction}). Details about
the utilized spectral range (shorter than the nominal one), resolving power,
exposure time, airmass and utilized spectrophotometric standard for each
pointing and configuration are shown in Table~\ref{log_observaciones}.

\begin{table*}
\centering
      \caption[]{Observation log \label{log_observaciones}}
              \begin{tabular}{ccccccccccccccccc}
            \hline
            \noalign{\smallskip}
Pointing & Grism &  Utilized spectral & Resolution & t$_{\mathrm{exp}}^{{\rm
(a)}}$ & Airmass & Seeing  & Standard \\ 
         &  &  range (\AA)         &  &             (s) &       &
($^{\prime\prime}$)   & star\\
            \noalign{\smallskip}
            \hline
            \noalign{\smallskip}
P1       & HR\_Orange & 5\,400--7\,200 & 3\,100 & $1(4\times425\,Obj + 120\,Bck)$ & 
1.05--1.09 & 0.4--0.8 & Feige~110\\
         & HR\_Blue   & 4\,200--5\,900 & 2\,550 & $2(4\times425\,Obj + 120\,Bck)$ & 
1.04--1.11 & 0.8--0.9 & EG~21\\
P2       & HR\_Orange & 5\,400--7\,200 & 3\,100 & $1(4\times425\,Obj + 120\,Bck)$ & 
1.04--1.06 & 0.6--0.8 & EG~21\\
         & HR\_Blue   & 4\,200--5\,900 & 2\,550 & $2(4\times425\,Obj + 120\,Bck)$ & 
1.04--1.07 & 0.7--1.1 & EG~21\\
            \noalign{\smallskip}
            \hline
         \end{tabular}
\begin{flushleft}
Notes:
$^{{\rm (a)}}$ Utilized notation: $N_{\mathrm{dither\,patterns}}
(N_{\mathrm{individual\,exp.}}  \times t_{\mathrm{exp}}$ Object $+
t_{\mathrm{exp}}$\,Background);
\end{flushleft}
\label{tabobslog}

\end{table*}

   \begin{figure*}[!th]
   \centering
\includegraphics[angle=0,width=0.38\textwidth, clip=]{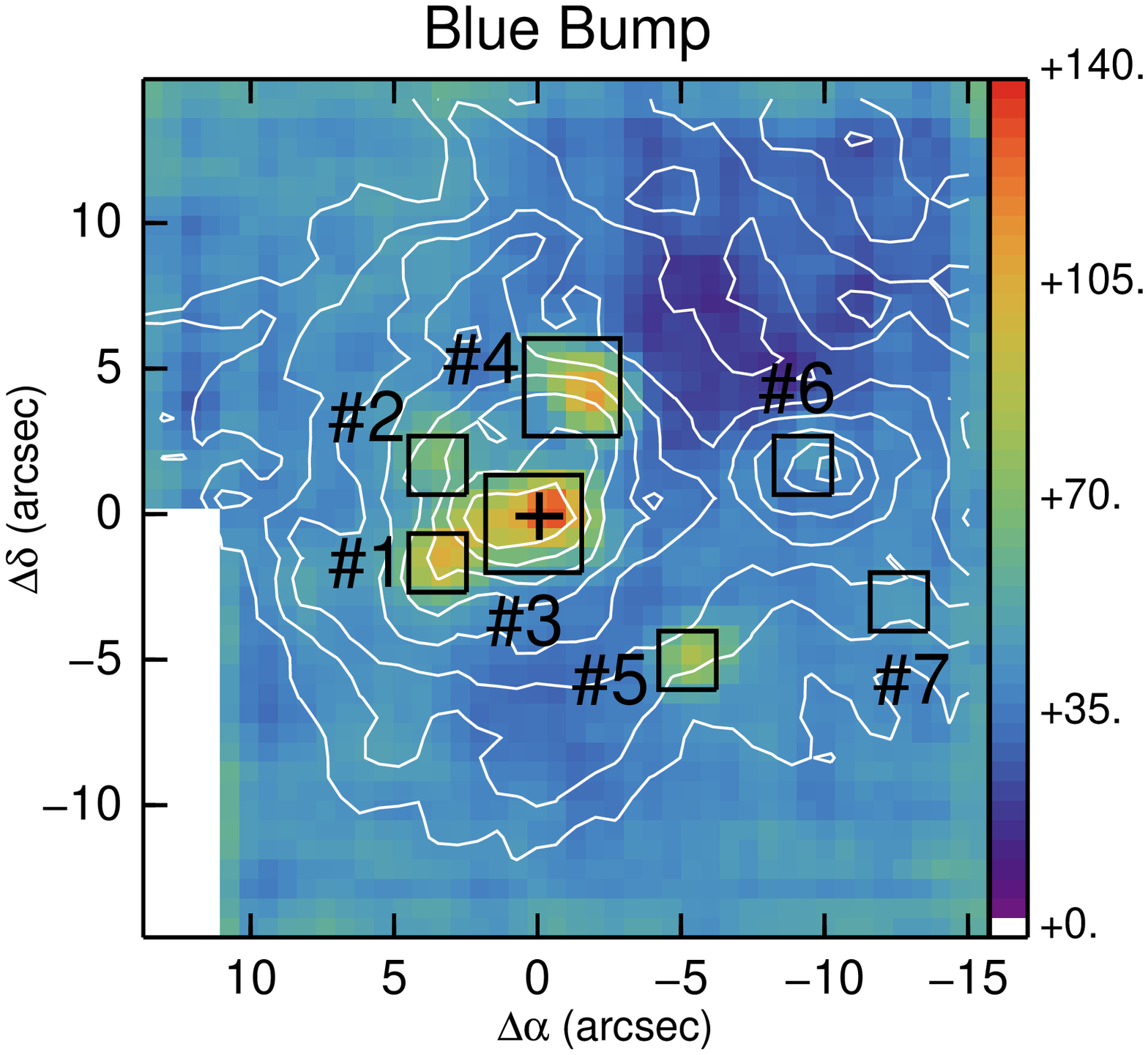}
\includegraphics[angle=0,width=0.38\textwidth, clip=]{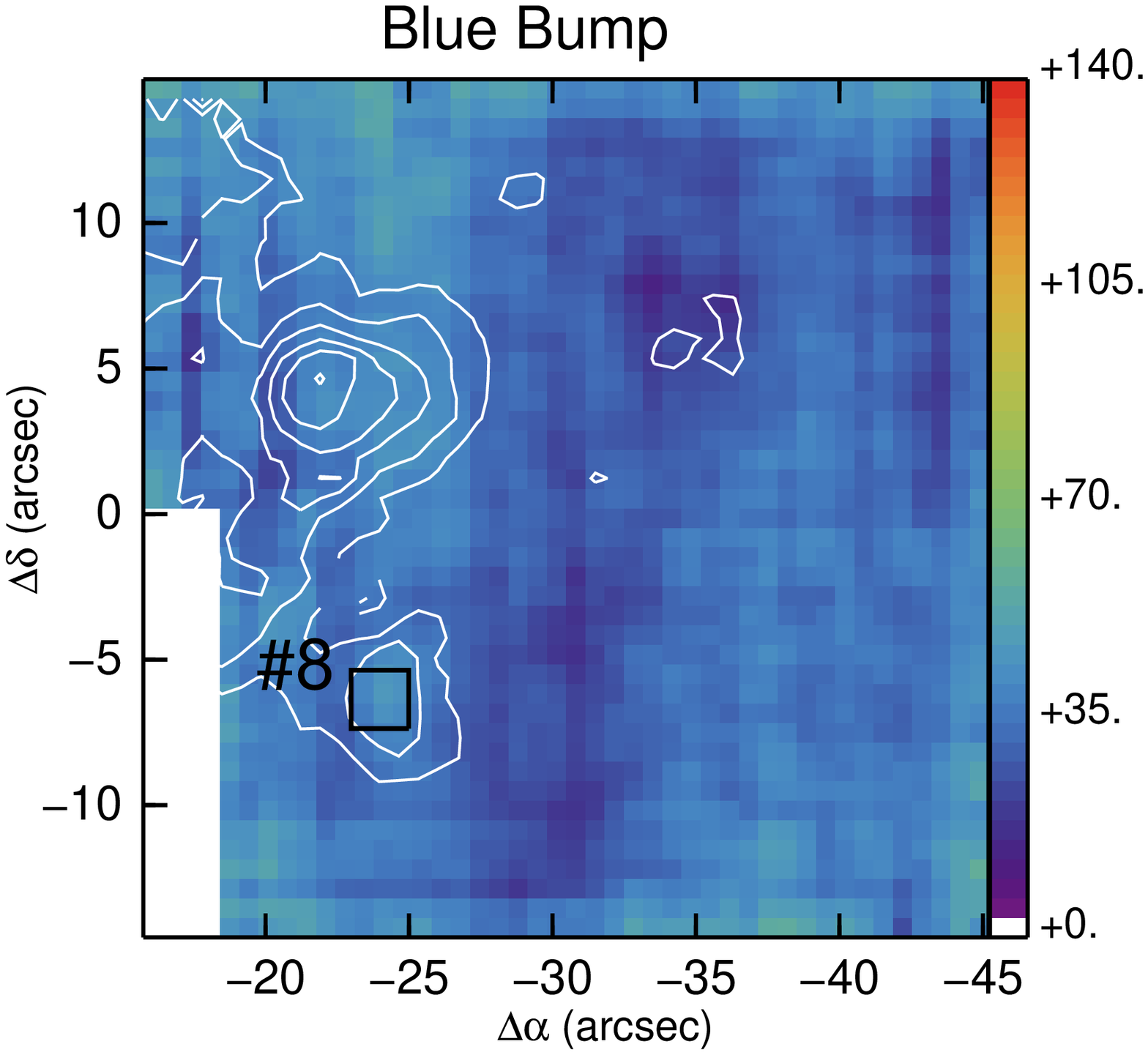}
\includegraphics[angle=0,width=0.38\textwidth, clip=]{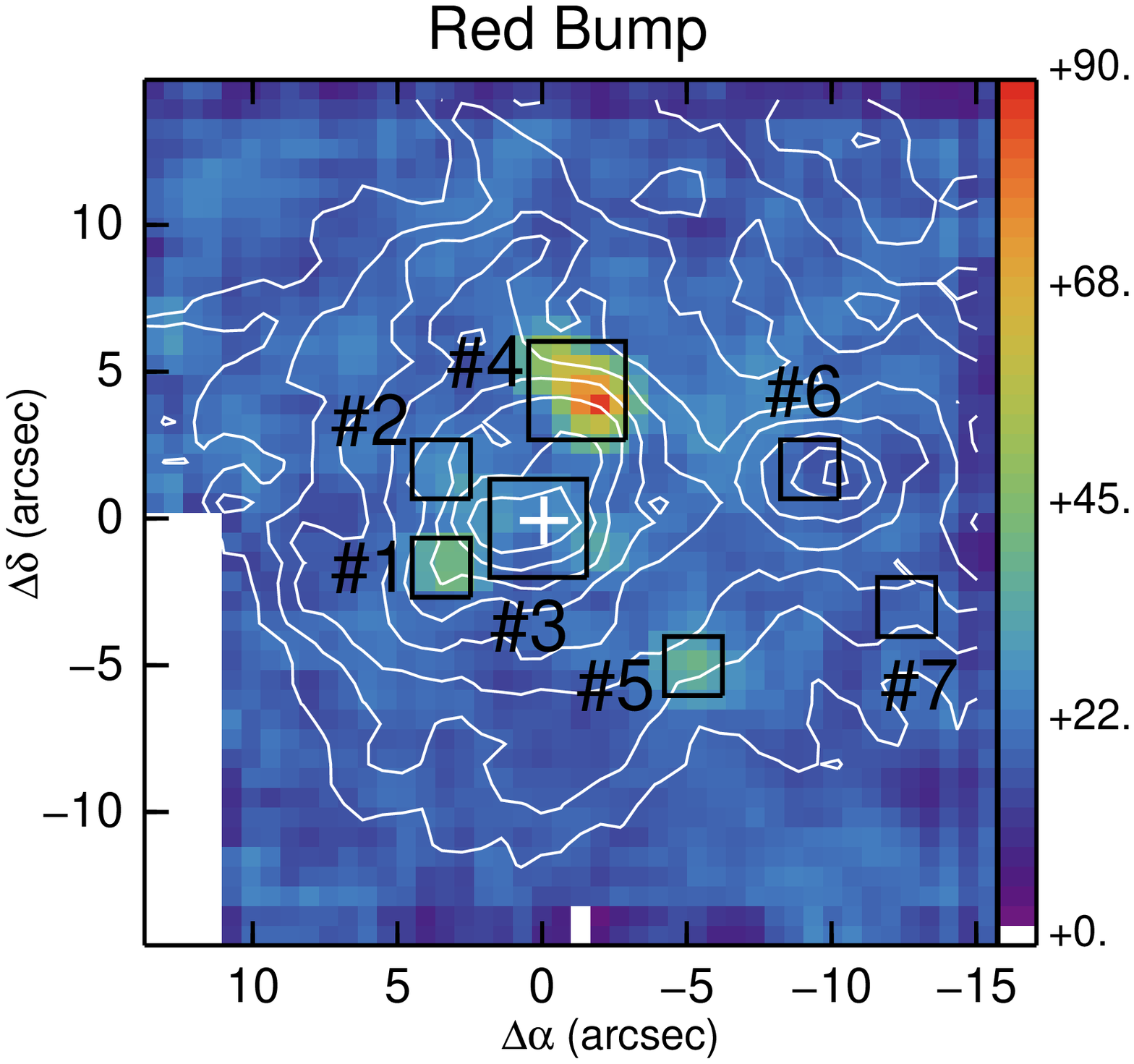}
\includegraphics[angle=0,width=0.38\textwidth, clip=]{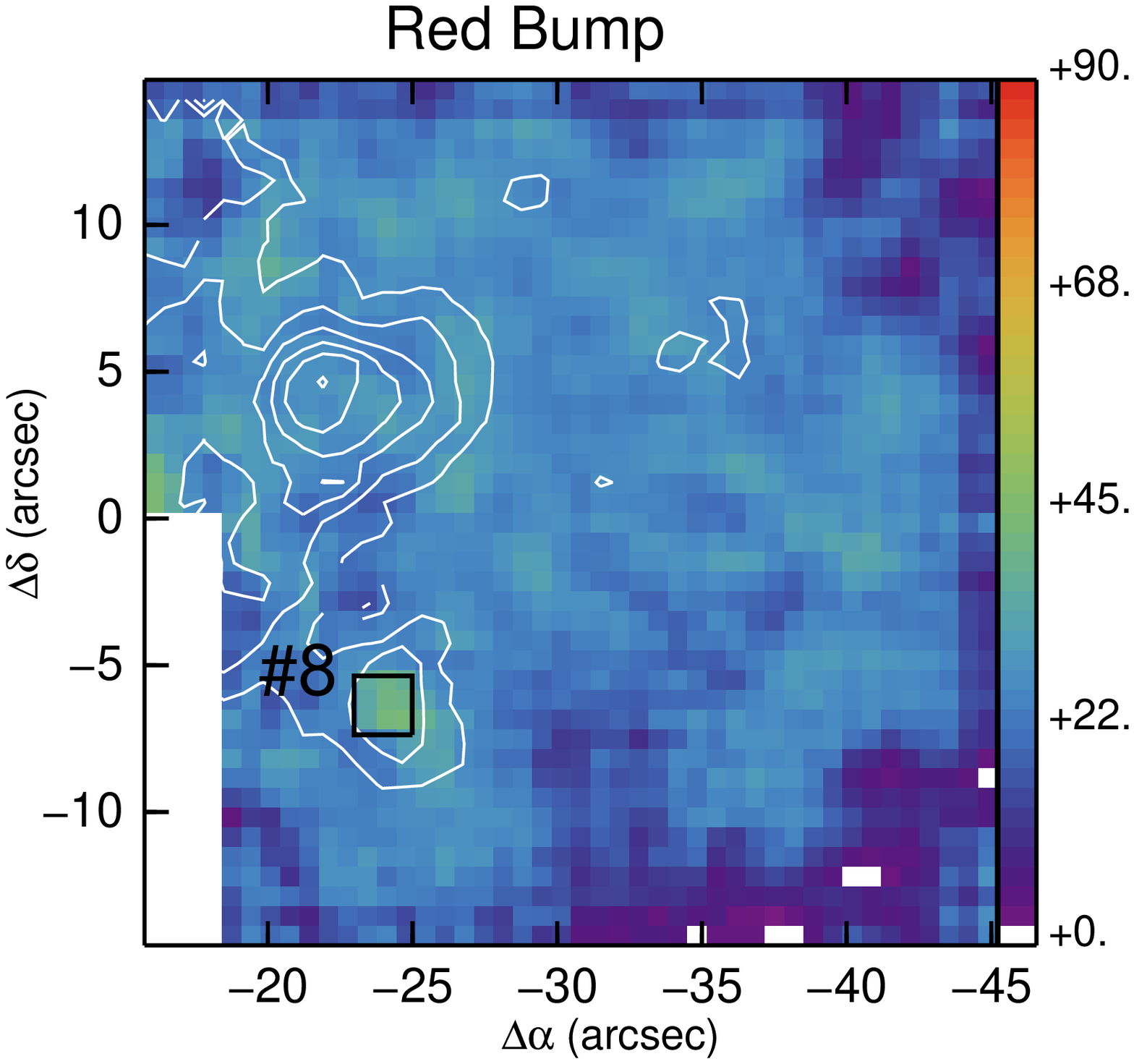}
\includegraphics[angle=0,width=0.38\textwidth, clip=]{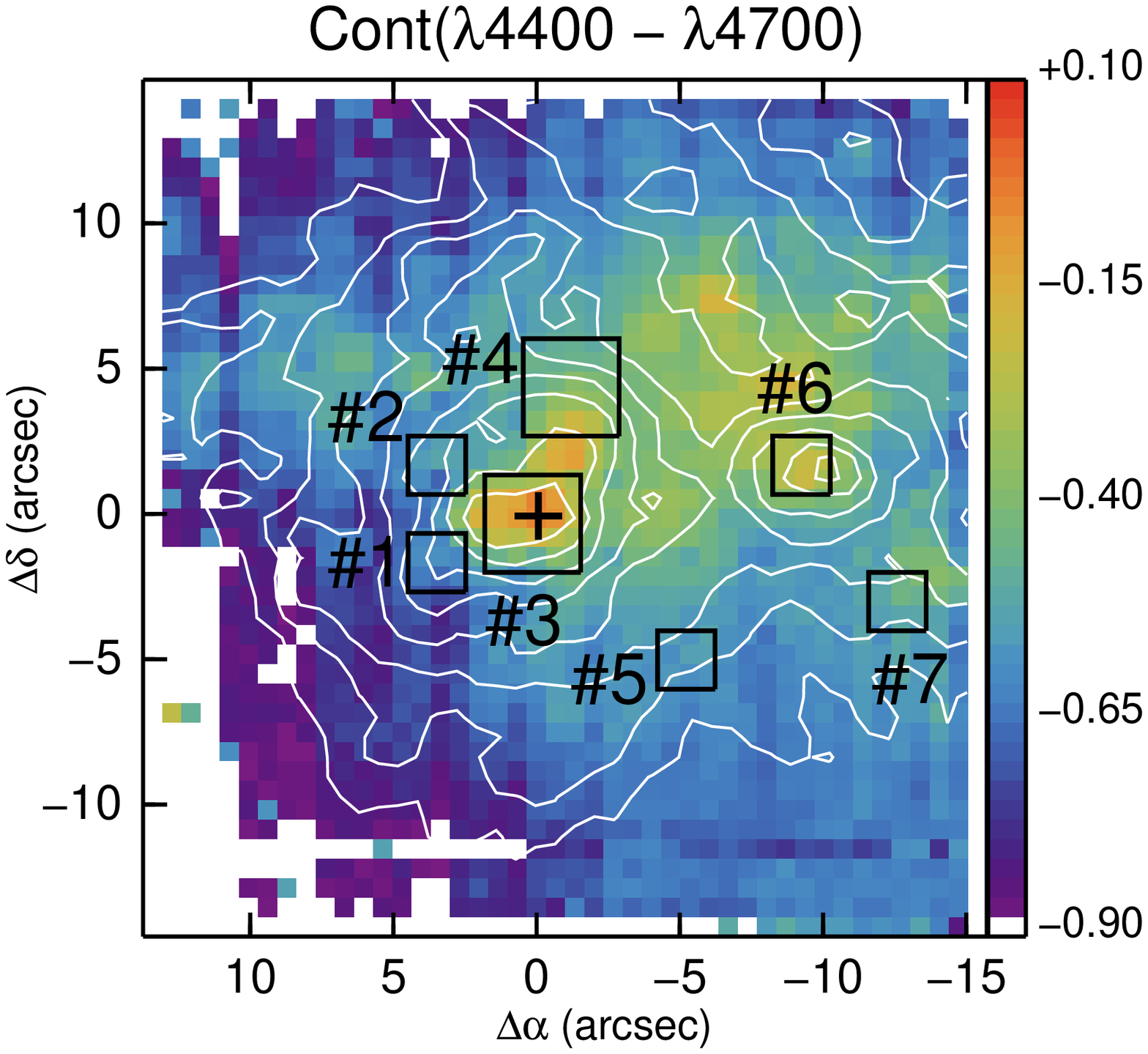}
\includegraphics[angle=0,width=0.38\textwidth, clip=]{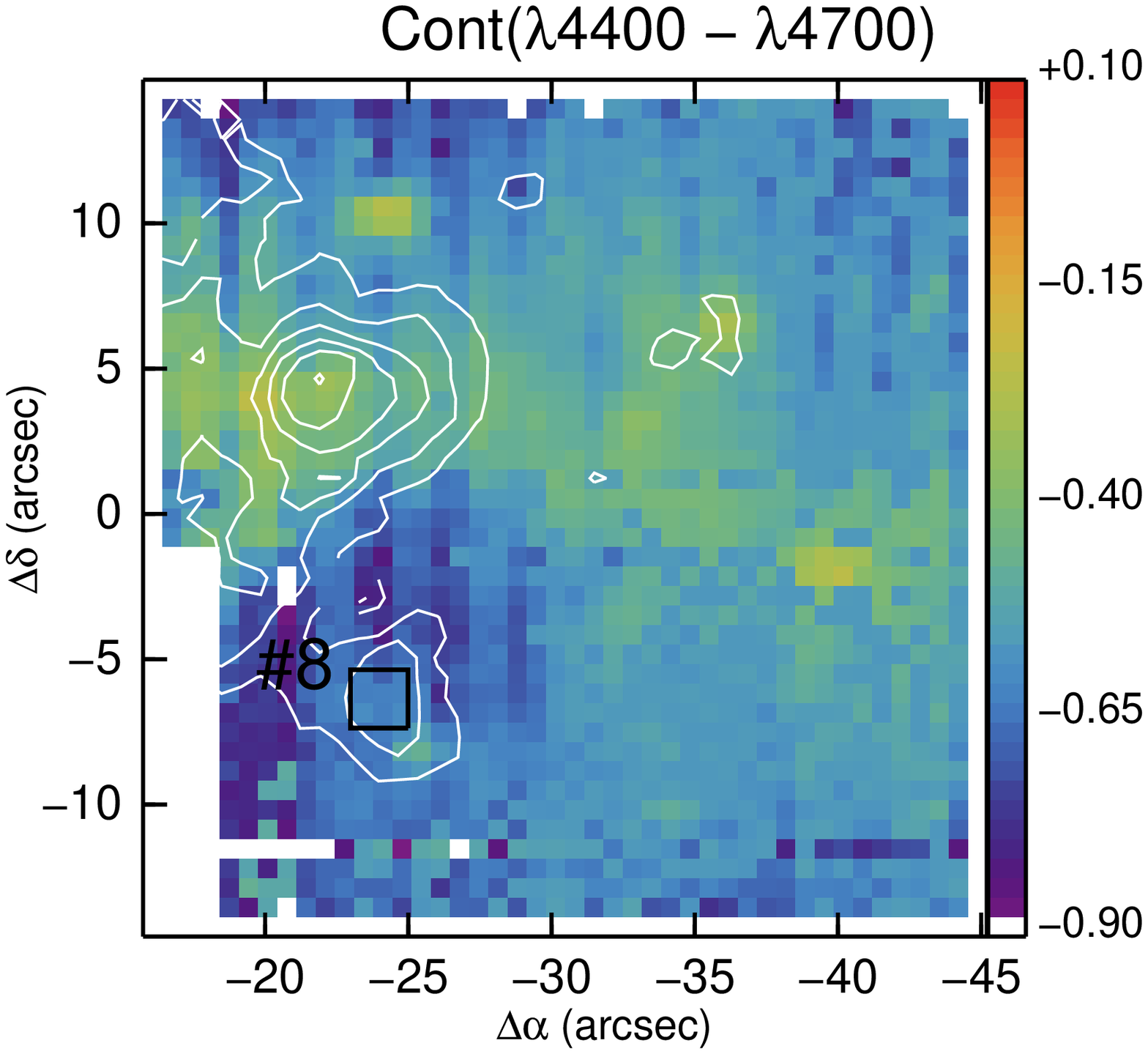}\\
   \caption[Maps for Wolf-Rayet features]{Maps showing the location of the
     Wolf-Rayet features with respect to the overall stellar population. Left
(right) column contains information for pointing 1 (2) (see Fig. \ref{apuntado2}).
\emph{Upper row:} Maps for the \emph{blue bump} emission.
\emph{Middle row:} Maps for the \emph{red bump} emission.
\emph{Lower row:} Maps for a emission line free stellar continuum.
%
Every map displays contours  tracing the observed \ha\ flux in steps of 0.3~dex
as derived from gaussian fitting (see Monreal-Ibero et al. in prep). 
Likewise, utilized squared apertures to extract the spectra at the locations
presenting W-R emission are shown and numbered in black.
A cross at the peak of emission in \ha\ marks our origin of coordinates.
The orientation is north up and east to the left.
The color bar on the right of each plot indicates the flux associated to the feature of interest in linear arbitrary units.
\label{mapswr}}
 \end{figure*}

\section{Data reduction \label{secreduction}}

Data were procesed using the {\pthreed} tool \citep[version
2.2,][]{san10,san11,san12}\footnote{All papers are freely available at the
{\pthreed} project web site \texttt{\href{http://p3d.sf.net}{http://p3d.sf.net}}.}
and some IRAF\footnote{The Image
  Reduction and Analysis Facility \emph{IRAF} is 
  distributed by the National Optical Astronomy Observatories which is
  operated by the association of Universities for Research in
  Astronomy, Inc. under cooperative agreement with the National
  Science Foundation.} routines.
The {\pthreed} tool  was utilized in the reduction of the individual exposures.
VIMOS data are provided in four files per detector that are reduced separately,
and combined after the flux calibration, to create a file containing the full
field with all 1600 spectra. Here is a description of the method.

As first steps, bias was subtracted and spectra were traced using the provided standard calibration files.
Then, a dispersion mask per observing block was created using the
available arc image.
The residual $r$ between the fitted wavelength
and the known wavelengths of the arc lines, was always in the range
$0.02\!\la\!r\!\la\!0.06\,\text{\AA}$.

In a fourth step, cosmic-ray hits are removed in the science images using the \textsc{PyCosmic} algorithm \citep{hus12}.

The fifth step was the extraction of the spectra.
With VIMOS, calculated profiles have
to be offset due to differences in the instrument flexures between the three
continuum lamp exposures and the science exposures. New centre positions are
calculated (for working fibres) in the science images, which are first
median-filtered on the dispersion axis. One median value of the difference
between all old and new centre positions is calculated and added to the profile
positions. 
After that,
spectra are extracted using the multi-profile deconvolution method
\citep{sha10}.
Data of all four quadrants of either grism are set up to use the same wavelength
array.
The dispersion mask was then applied in a sixth step.

To estimate the accuracy achieved in the wavelength calibration for the
science frames we fitted 
 the \oi$\lambda$6300 sky line in each spectrum by a Gaussian for the
data taken with the HR\_Orange grism. 
The standard deviation of the distribution of centroids of the lines was
$\sim0.02$~\AA. Assuming this is a proxy of the quality in the wavelength
calibration, this implies an accuracy of $\sim$1~km s$^{-1}$.
For the HR\_Blue, the \oi$\lambda$5577 was used. We also measured a standard
deviation of $\sim0.02$~\AA. 

Similarly, the width of the gaussian can be used as a proxy of the 
spectral resolution. We measured a  instrumental width 
$\sigma\sim0.75(\pm0.06)$~\AA\ and $\sim0.69(\pm0.05)$~\AA\ using the
\oi$\lambda$6300 and \oi$\lambda$5577 lines, for the HR\_Orange and HR\_Blue
grism configurations, respectively. This translates to $\sigma_{instru}
\sim$36~km~s$^{-1}$ for both configurations. 

In a seventh step, a correction to the fibre-to-fibre sensitivity variations was
applied. 
For that,
the spectra were divided by the mean spectrum of an extracted and normalised
continuum lamp flat-field image made out of the combination of three
continuum-lamp exposures, excluding all spectra of broken and low-transmission
fibres.
Then, extracted spectra were flux calibrated using the
standard-star exposure of the respective night and grism. 

The last step of the reduction of the individual exposures was the creation of a
data cube  by combining the 
flux calibrated images of the four separate detectors and
reordering the individual spectra according to their position within the
VIMOS-IFU.

After this reduction, measured fluxes of the telluric lines still showed some
variations depending on both the detector and the spatial elements \emph{per se} \citep[also
see][]{lag12b}. All spectra within a datacube are therefore normalised to
achieve the same integrated median flux in a telluric line. For that, we used
the brightest observed line in each configuration (i.e.  $\lambda\!=\!5577\AA$
and $\lambda\!=\!6300\AA$ with data of both HR$\_$Blue and HR$\_$Orange,
respectively).

{\pthreed} evaluates the offsets due to differential atmospheric refraction
(DAR) using the expressions listed by \citet{cid96} and offers the possibility
of correcting for it as described in \citet{san12}.
In our specific case, these were always $\lsim$0.2~spaxels and
$\lsim$0.3~spaxels for the utilized spectral range in the HR\_Orange and
HR\_Blue configuration, respectively. Since this is relatively small and a
correction for differential atmospheric refraction would imply an extra
interpolation in the data, we took the decision of not applying such a
correction. 

After the cubes for the individual exposures were fully reduced, all science
cubes in each observing block were combined by using the offsets commanded to
the telescope and the IRAF task \texttt{imcombine}. Then a high signal-to-noise
background spectrum was created by combining the individual spectra of the
background exposure and subtracted afterwards. Finally, the two resulting cubes
for each HR$\_$Blue pointing were combined.

\section{Results \label{secresults}}

\subsection{Pinpointing Wolf-Rayet emission \label{subsecpinpointing}}

\begin{table}
\caption{Rest-frame wavelength of the narrow filters proposed by \citet{bri08}.}       
\label{narrowfilters}      
\centering          
\begin{tabular}{lccc}     
\hline\hline       
Filter &
$\lambda_{cen}$(\AA)   &
$\Delta\lambda$(\AA)\\
\hline                    
Blue Bump                  & 4705 & 100\\  
Blue Bump Continuum & 4517 & 50\\
                                   & 4785 & 50\\
Red Bump                   &  5810 & 100\\
Red Bump Continumm & 5675 & 50\\
                                     &  5945 & 50\\
\hline                  
\end{tabular}
\end{table}
%
%

To localize the W-R emission in NGC~625, narrow \emph{tunable
filters}  were simulated using the bands
proposed in \citet{bri08} -  listed in Tab. \ref{narrowfilters} for completness - and the redshift of the galaxy.
%
Depending on the characteristics of the stellar continuum, this very simple
procedure may create flux maps with
negative values at specific spatial elements. To avoid this, we added a common
flux offset determined \emph{ad hoc} to
all spatial elements.
Note that this does not affect our purposes (i.e. identifying  the
locations with emission excess at the bumps). 
Then, following the methodology explained in \citet{mon11}, 
we convolved the resulting image with a  Gaussian of $\sigma$=0\farcs7
 in order
to better identify the places that might present W-R emission.
The resulting maps for the \emph{blue} and \emph{red} bumps are presented in
the first and second rows of Fig. \ref{mapswr}.
We set our origin of coordinates at the peak of emission in \ha.
Additionally, a map to trace the overall stellar structure, made by simulating the action of a relatively broad filter (4400-4700\, \AA), de-contaminated from any emission feature (see Paper II, in prep), is presented in the third row. 
Likewise, to have a reference for the structure of the ionised gas, contours corresponding to the \ha\, emission are overplot in each map.
In short, these maps show that: i) W-R emission in this galaxy is extended and
with peaks of emission in multiple
locations; ii) the peaks for the \emph{blue bump} emission may or may not coincide with
those of the \emph{red bump} emission;
iii) these peaks, in general, do not necessarily coincide with either  the overall
general stellar or ionised gas emission distribution. Specifically,
the stellar distribution presents several peaks of emission. The most important
one is associated to the peak of
emission in \ha\ (contours in Fig. \ref{mapswr}) and with the largest peak for the \emph{blue bump} emission.
Additionally, the overall stellar distribution present some peaks at $\sim(-7\farcs0,5\farcs0)$ which are
associated to a plethora of young
(i.e. \lsim20~Myr) stars identified by means of HST broad band imaging
\citep{can03,mcq12}. However, no W-R emission in
any of the bumps is seen there.
In general, the \emph{blue bump} emission is
more concentrated than the \emph{red bump} emission.
Indeed the only W-R emission detected in P2 is in the red bump,  at about 400~pc from  the peak of
emission in \ha.

Those locations with eight\footnote{This would correspond to an area equivalent to that of a circle with $r\sim1\farcs1$, comparable to the seeing of these observations.}
or more adjacent spatial elements showing values 
0.8 times standard deviations above the median of the whole image in at least one of
the maps were identified as candidates with W-R emission whose spectra should be
extracted to be analyzed in more detail.
The eight locations satisfying this criterion, as well as the square apertures
that we used to extract the spectra, are
numbered and marked in every map of Fig. \ref{mapswr}. Their analysis will be
the subject of the next section.

   \begin{figure*}[!th]
   \centering
\includegraphics[angle=0,width=0.45\textwidth, clip=]{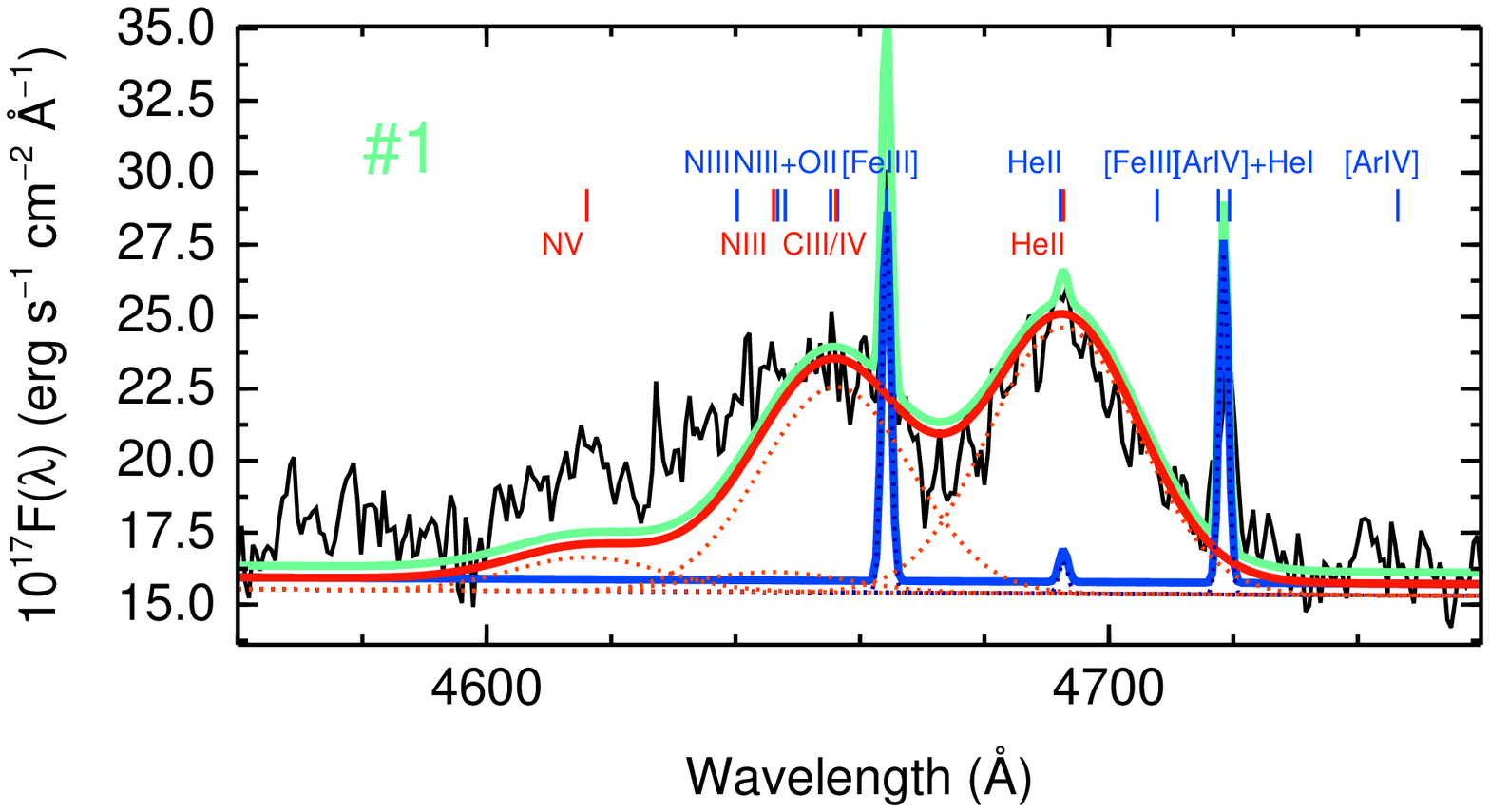}
\includegraphics[angle=0,width=0.45\textwidth, clip=]{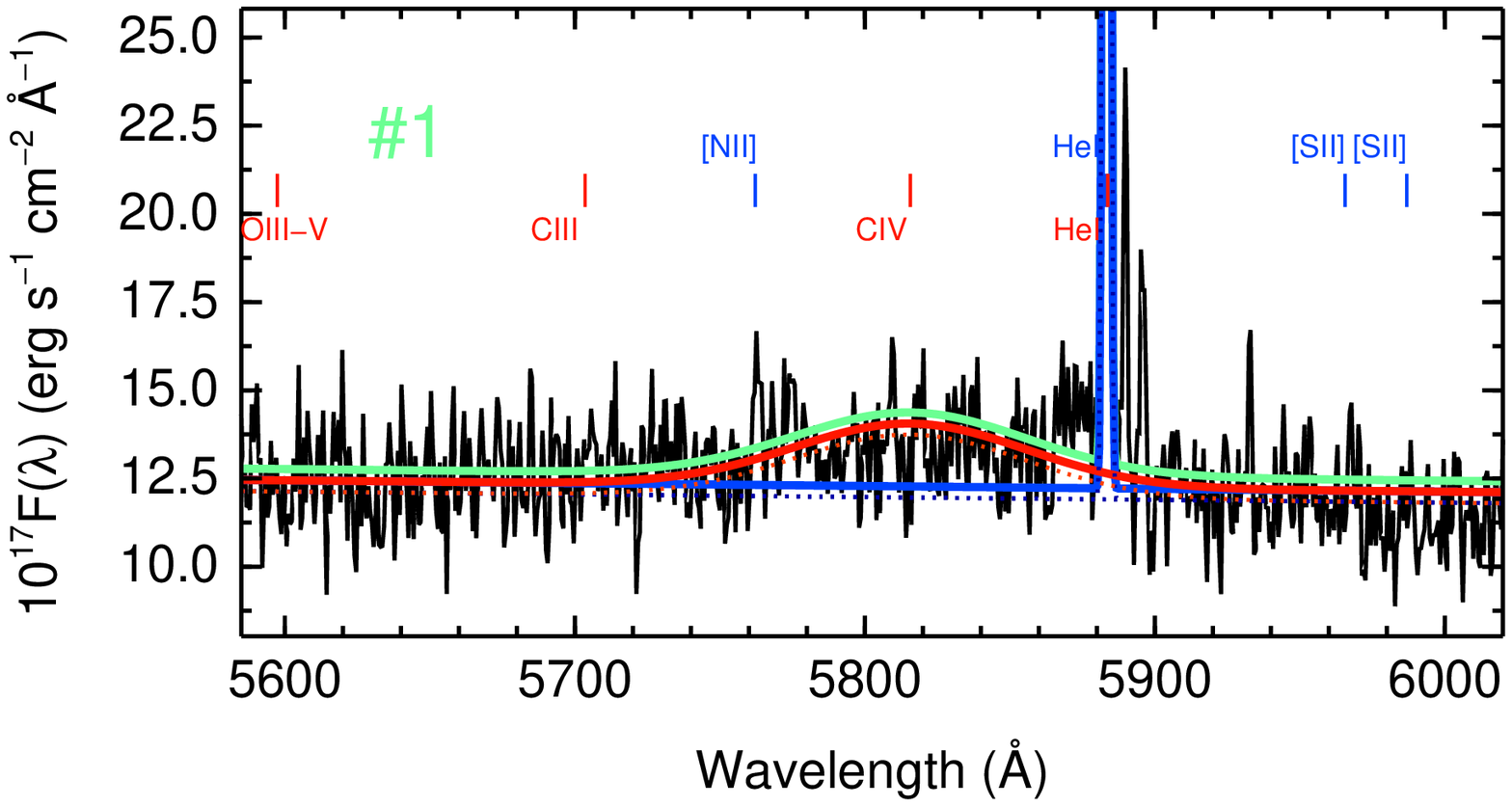}\\
\includegraphics[angle=0,width=0.45\textwidth, clip=]{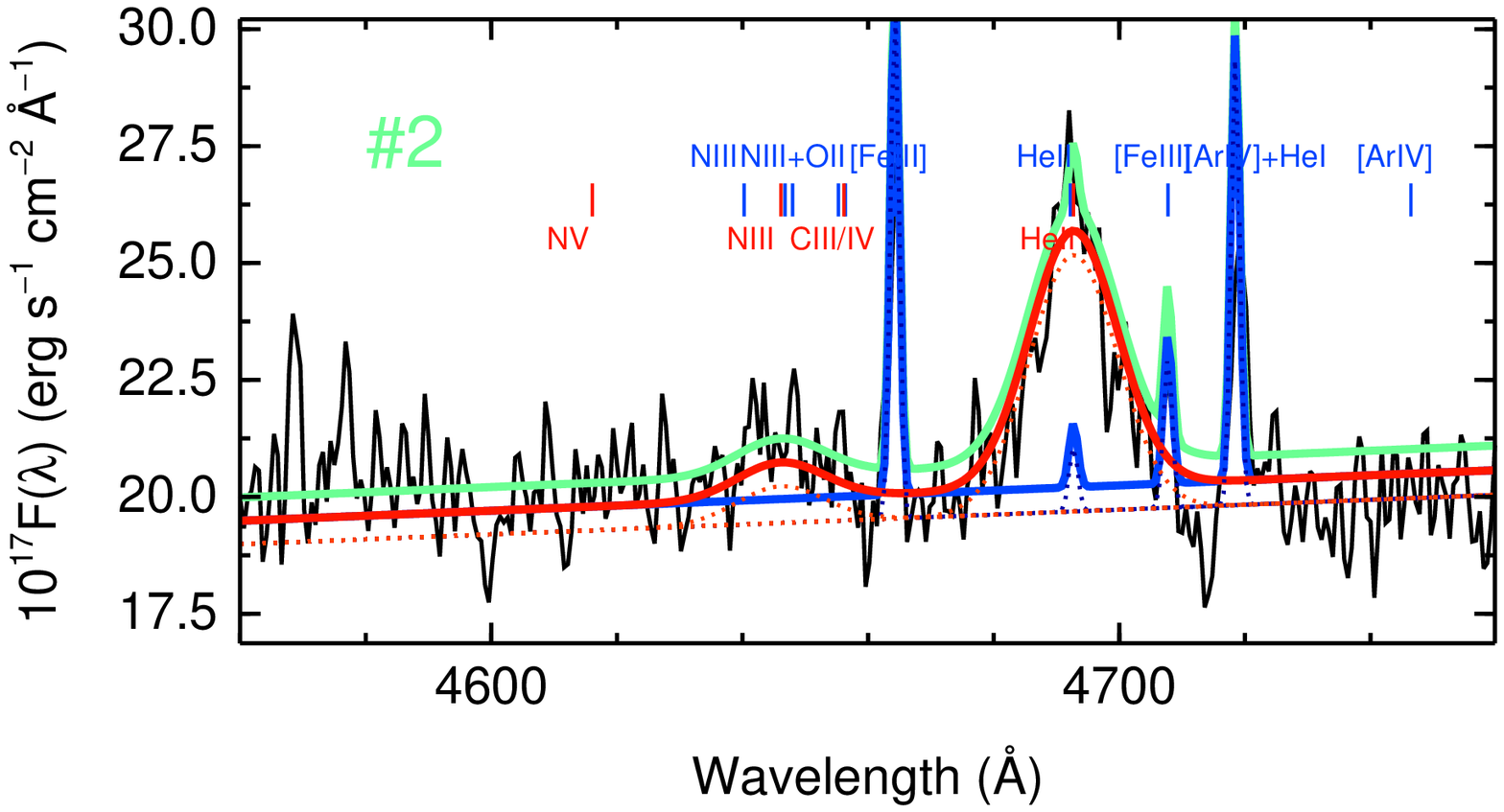}
\includegraphics[angle=0,width=0.45\textwidth, clip=]{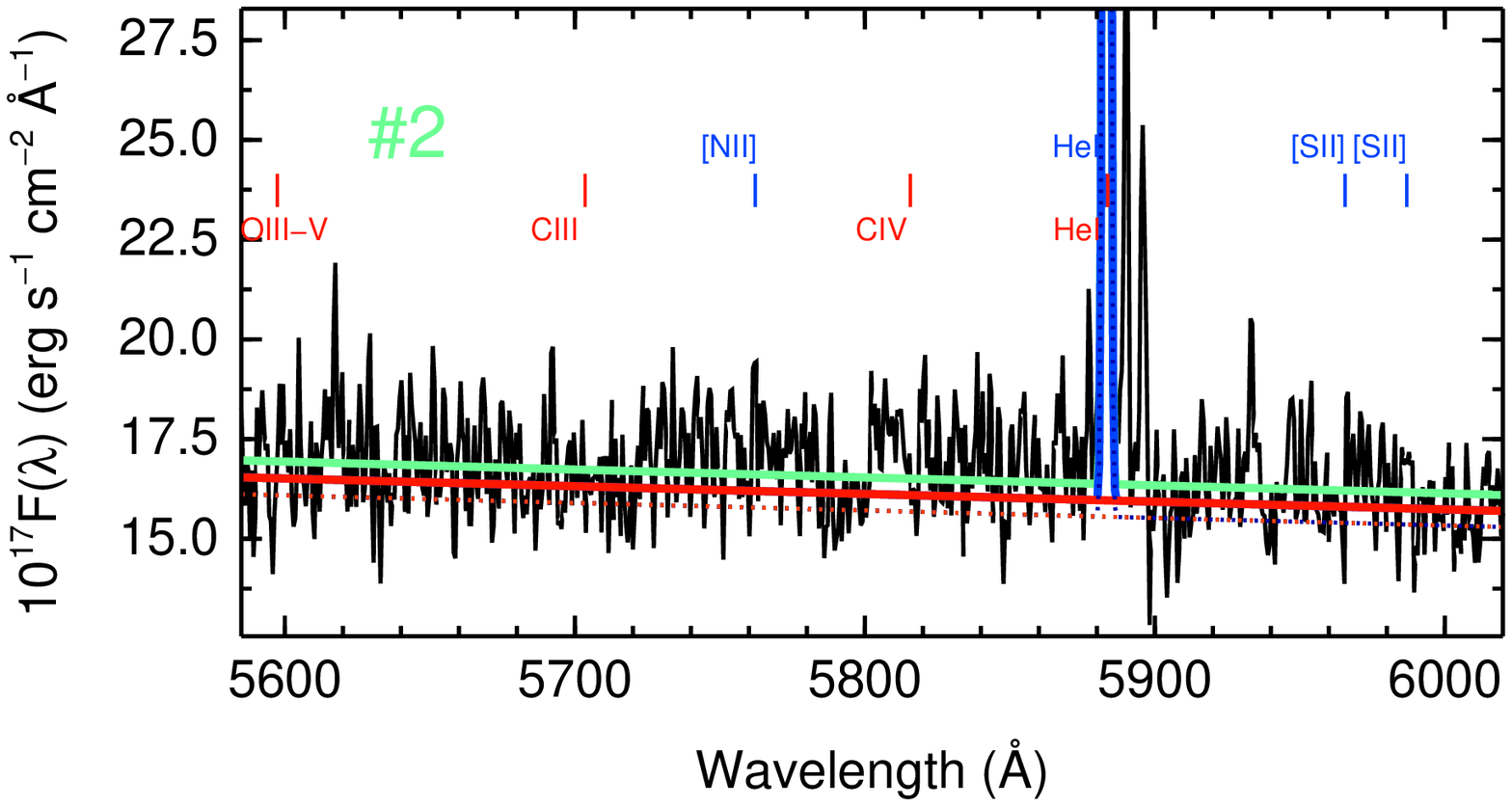}\\
\includegraphics[angle=0,width=0.45\textwidth, clip=]{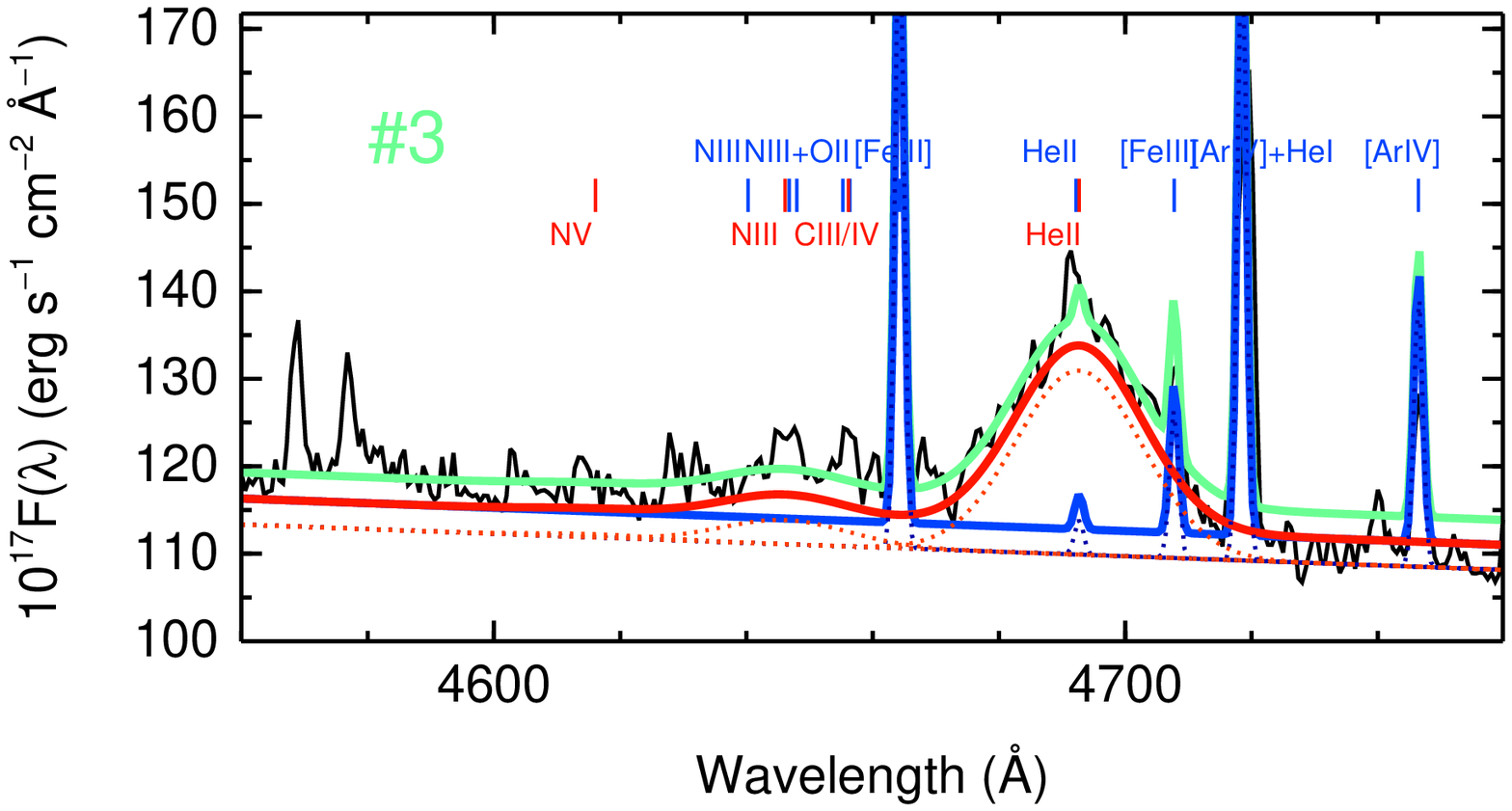}
\includegraphics[angle=0,width=0.45\textwidth, clip=]{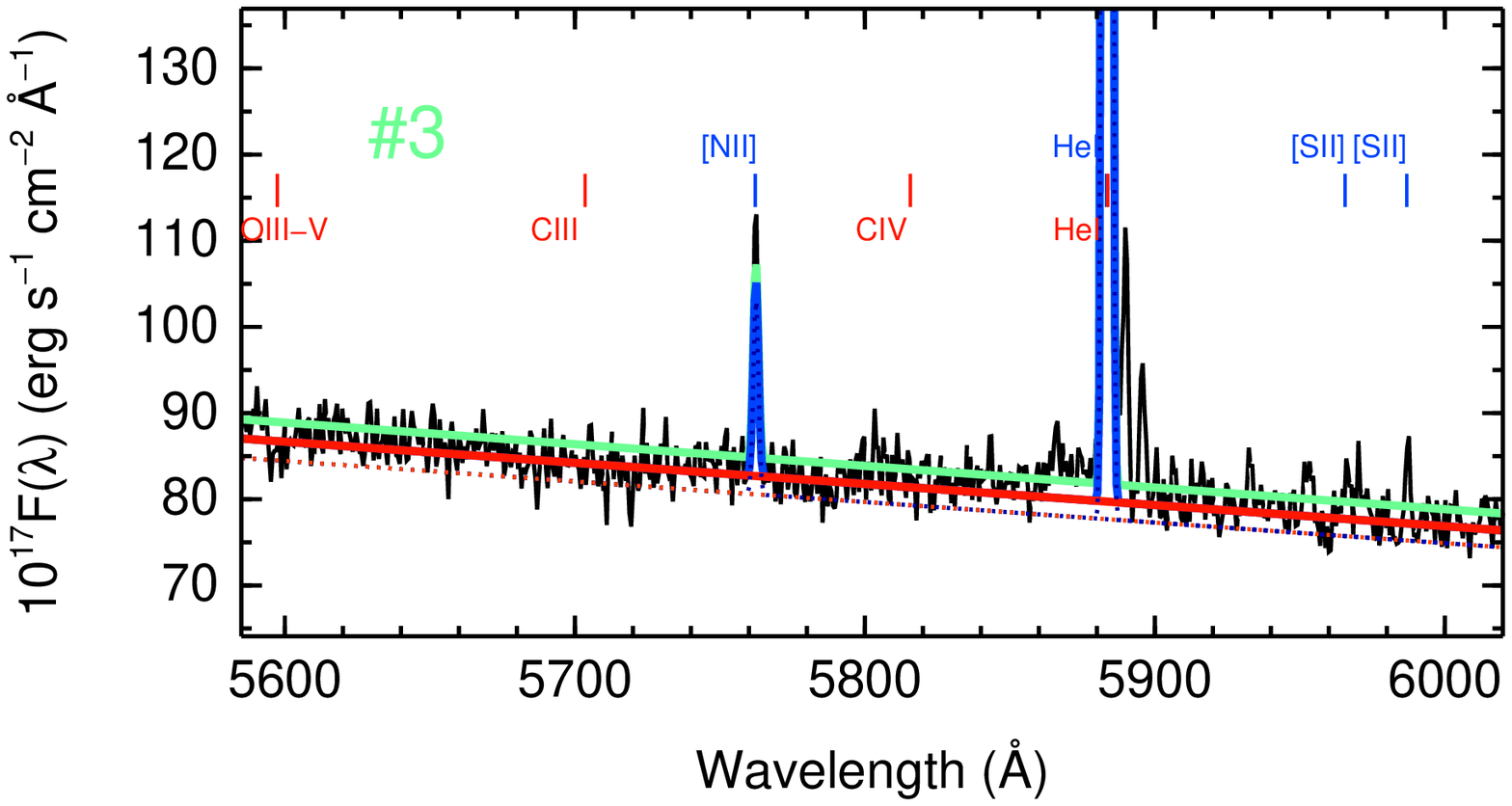}\\
\includegraphics[angle=0,width=0.45\textwidth, clip=]{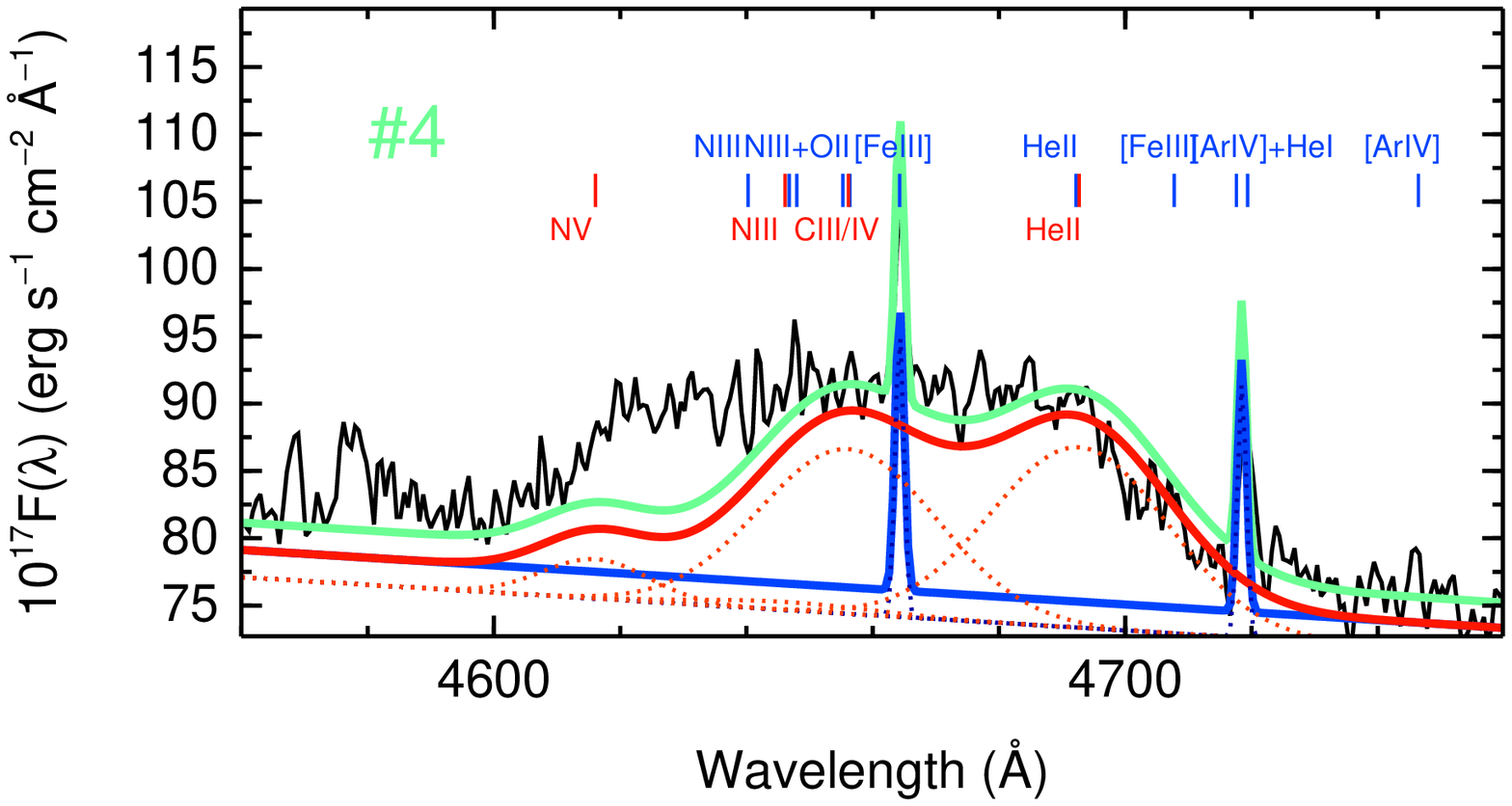}
\includegraphics[angle=0,width=0.45\textwidth, clip=]{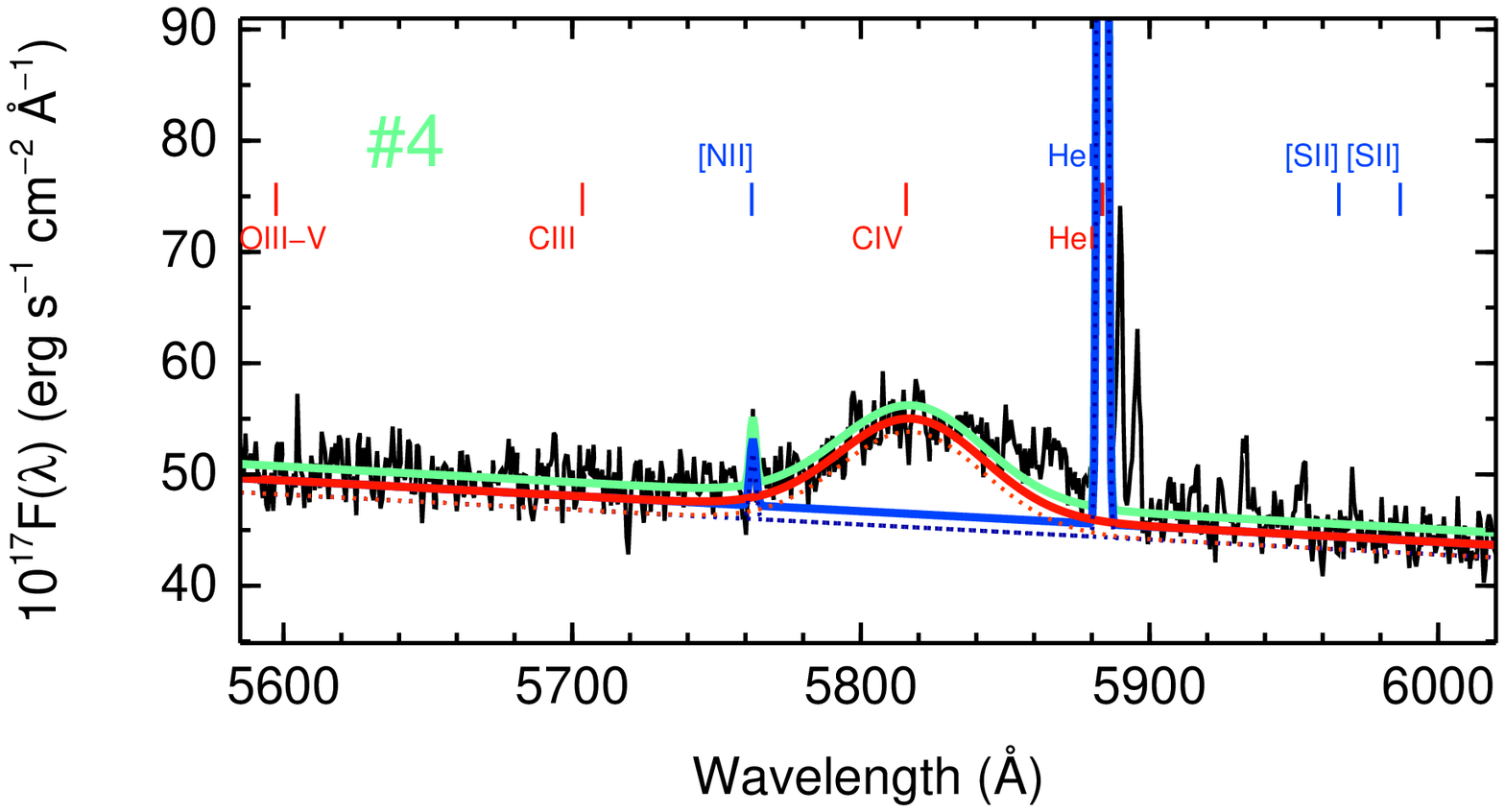}\\
   \caption[Spectra in the 6 selected apertures presenting Wolf-Rayet
features]{Spectra in selected apertures presenting Wolf-Rayet features.
The total modeled spectrum is shown in green whilst the fluxes corresponding to the WR stellar features and nebular lines are shown in red and blue continuous lines respectively, with a small offset. Additionally, individual fits for each feature are shown as dotted lines.  
The positions of the nebular emission lines are indicated with blue ticks and
labels, whilst those corresponding to Wolf-Rayet features appear in red. 
Apertures were numbered according to decreasing right ascension and increasing
declination. Their locations and sizes are shown in Fig. \ref{mapswr}.
 \label{spectrawr}}
 \end{figure*}

\begin{figure*}[!th]
   \centering
\includegraphics[angle=0,width=0.45\textwidth, clip=]{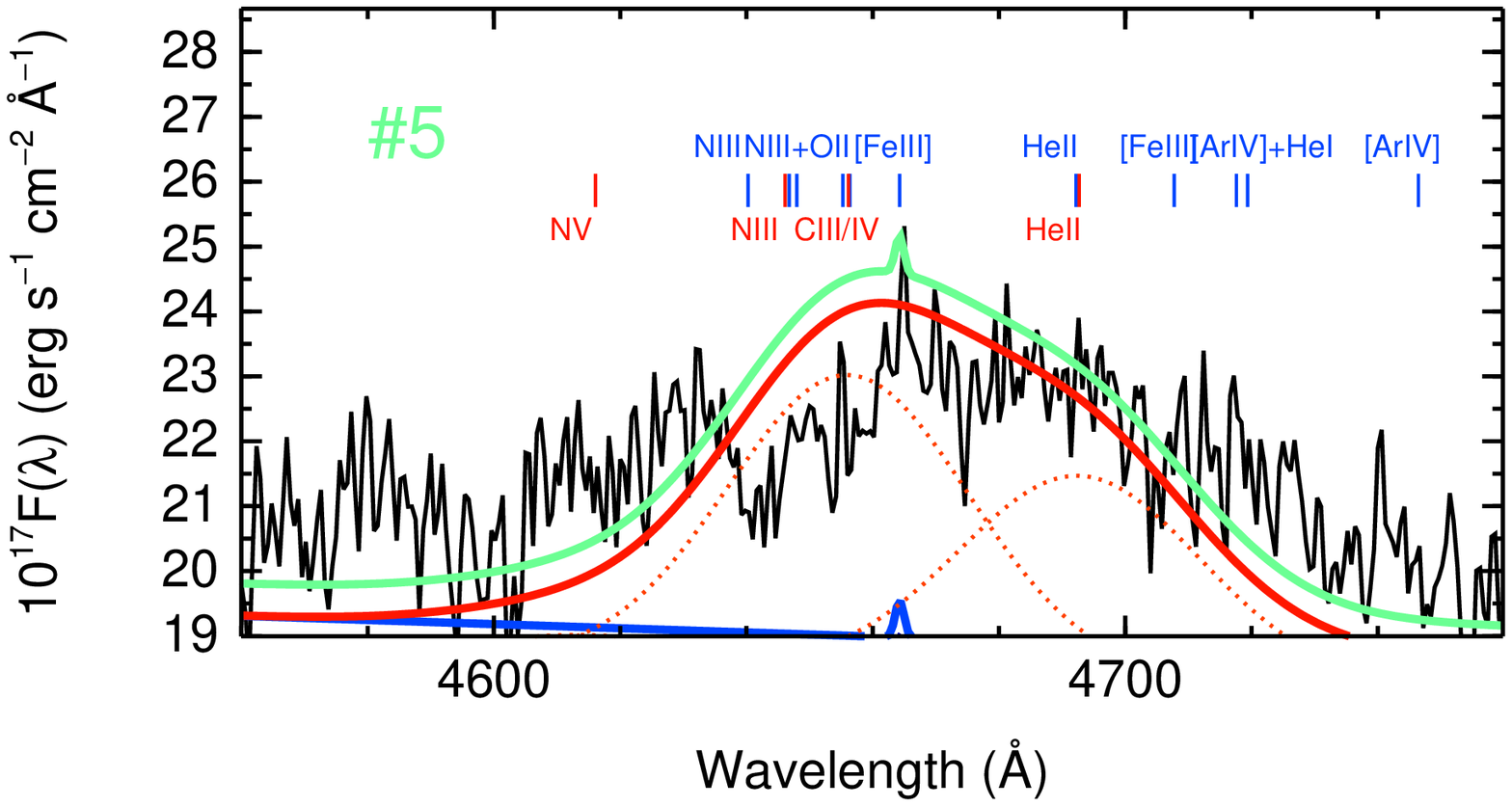}
\includegraphics[angle=0,width=0.45\textwidth, clip=]{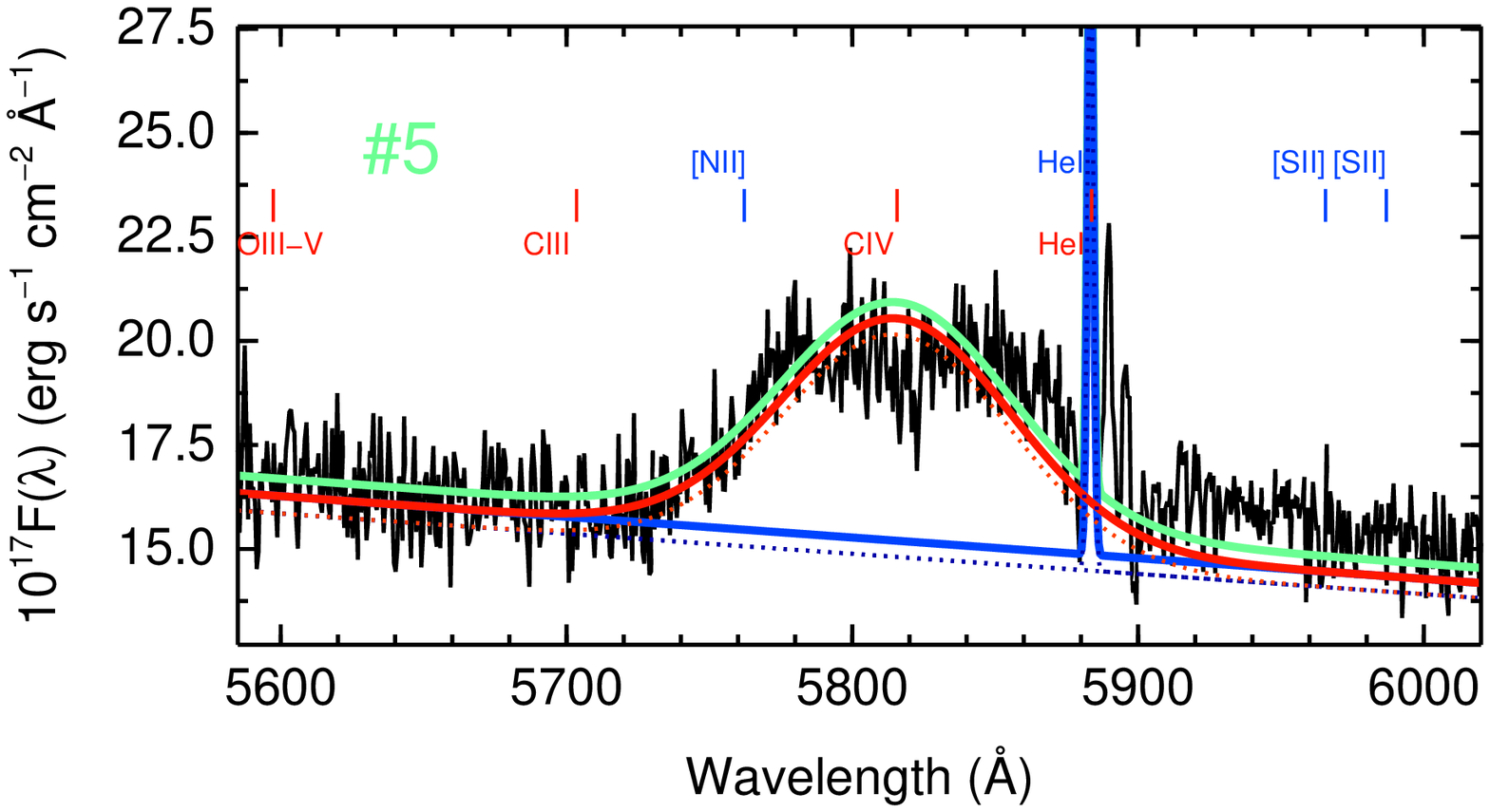}\\
\includegraphics[angle=0,width=0.45\textwidth, clip=]{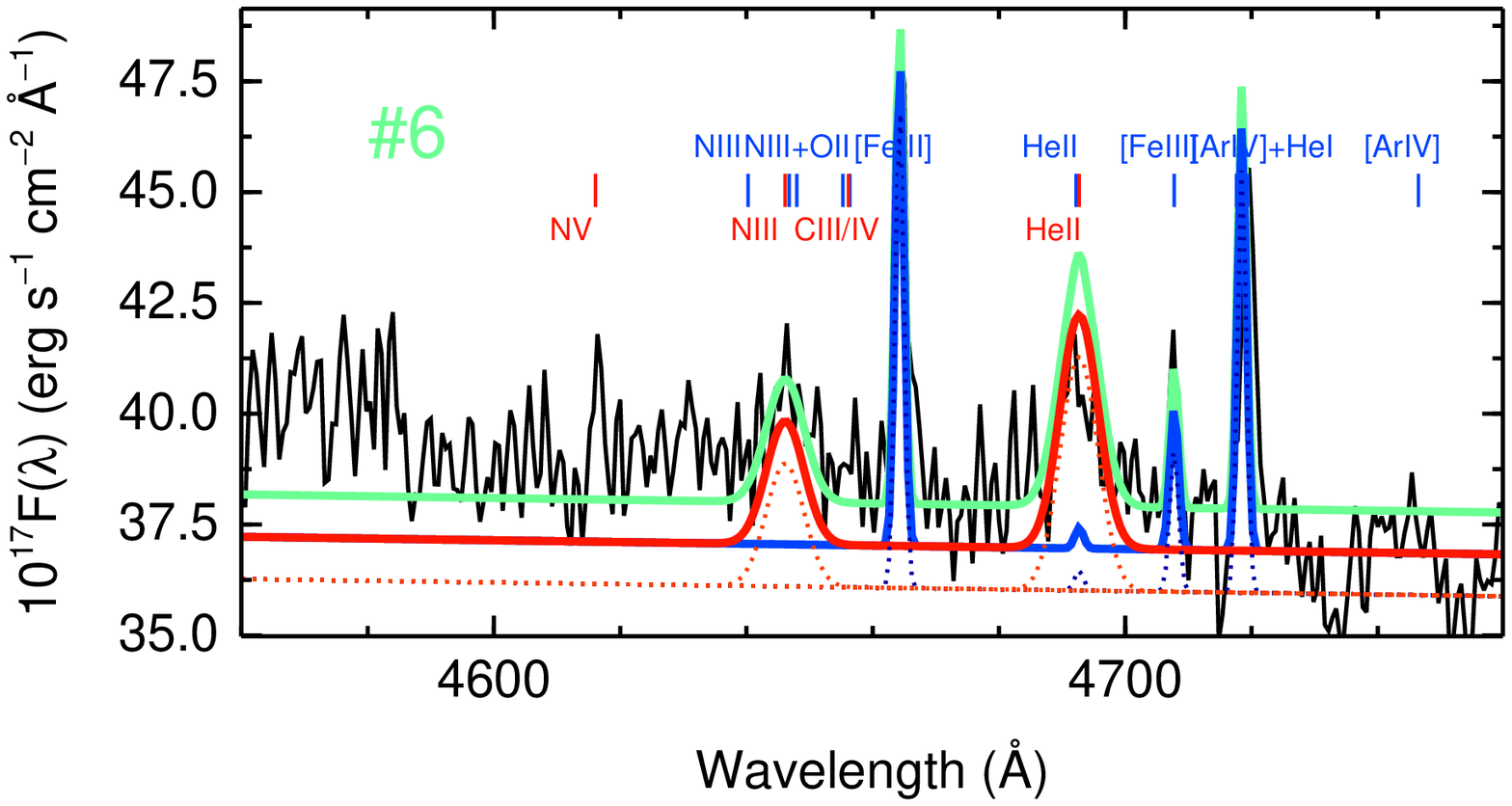}
\includegraphics[angle=0,width=0.45\textwidth, clip=]{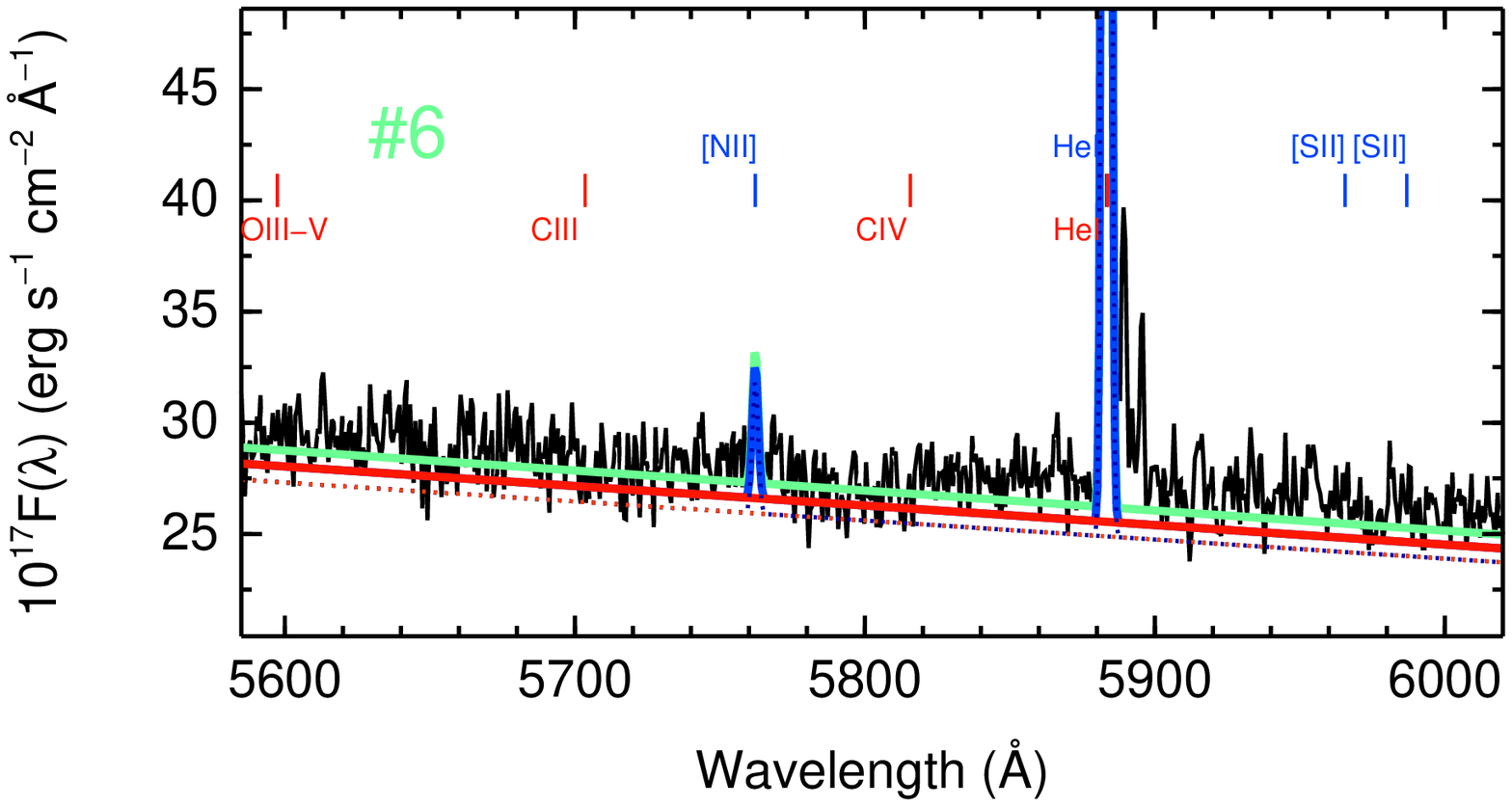}\\
\includegraphics[angle=0,width=0.45\textwidth, clip=]{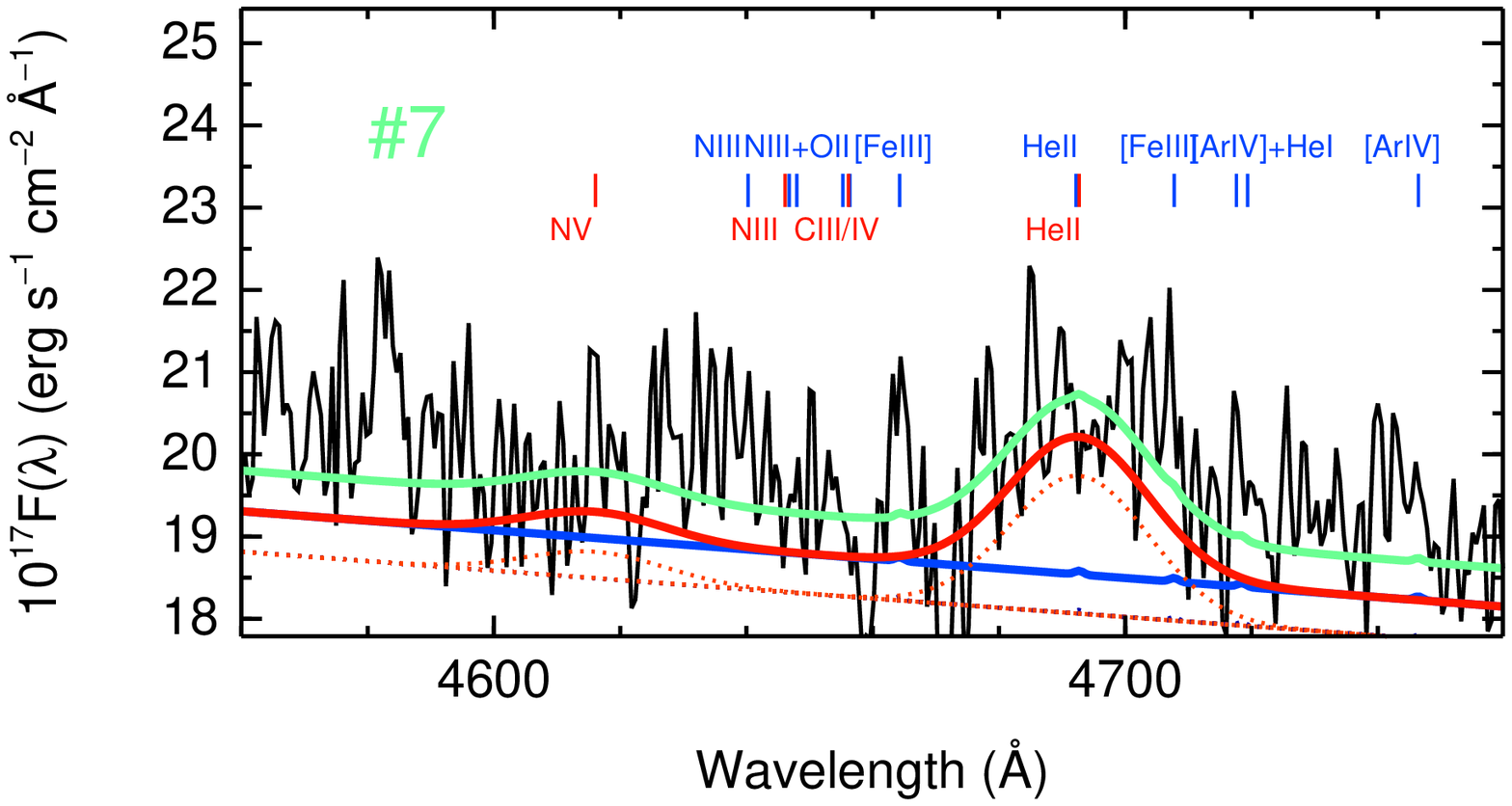}
\includegraphics[angle=0,width=0.45\textwidth, clip=]{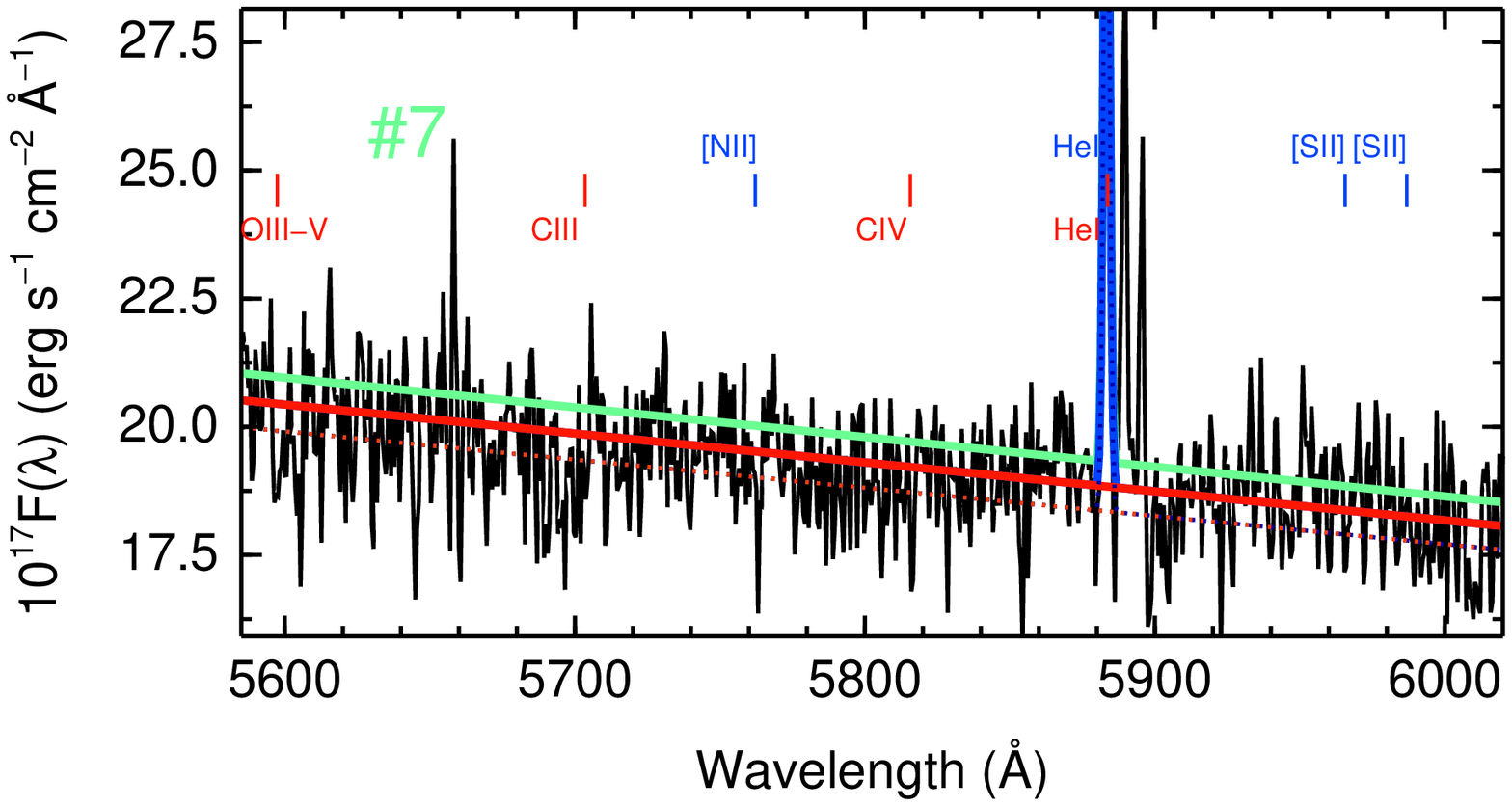}\\
\includegraphics[angle=0,width=0.45\textwidth, clip=]{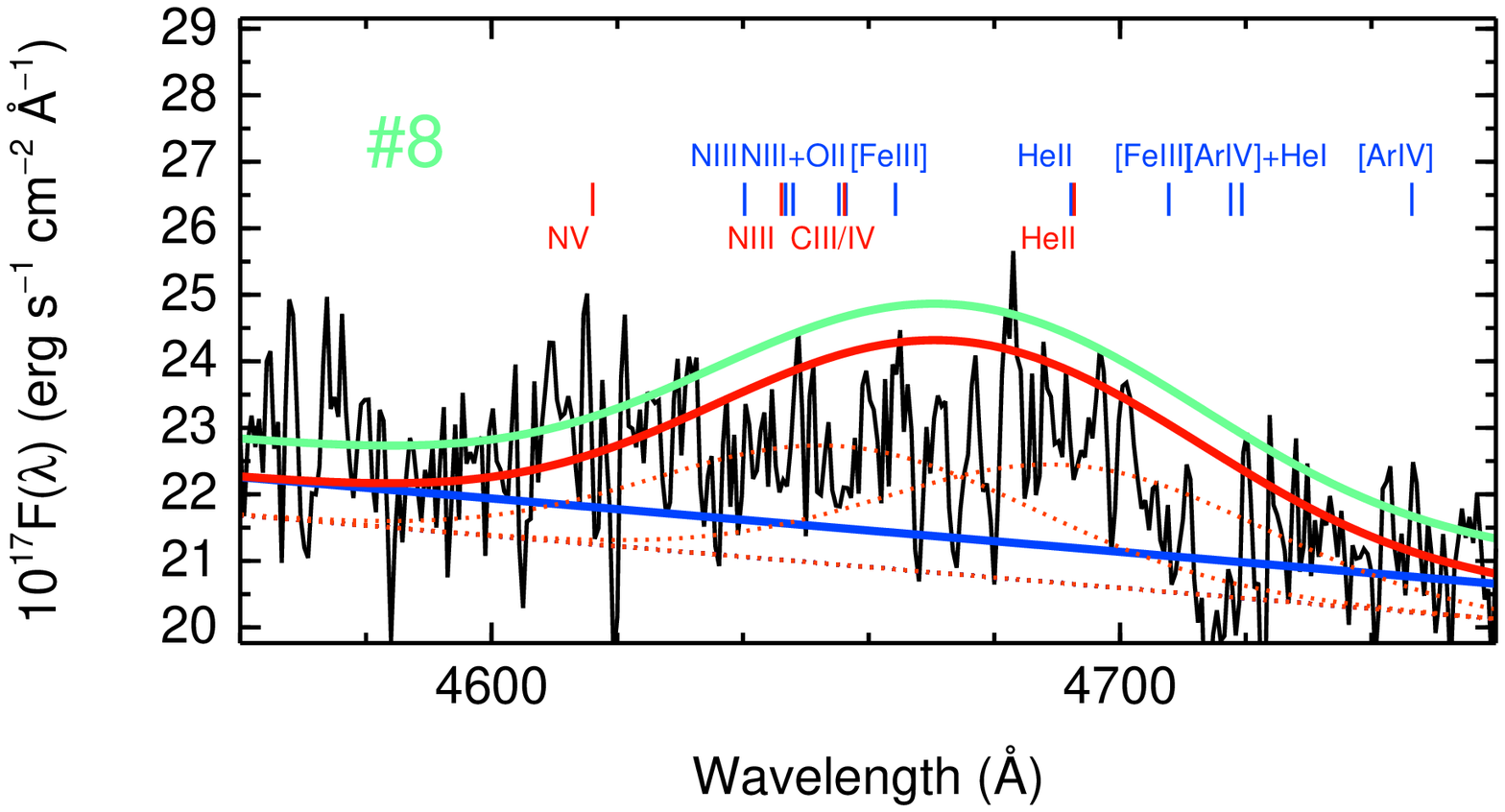}
\includegraphics[angle=0,width=0.45\textwidth, clip=]{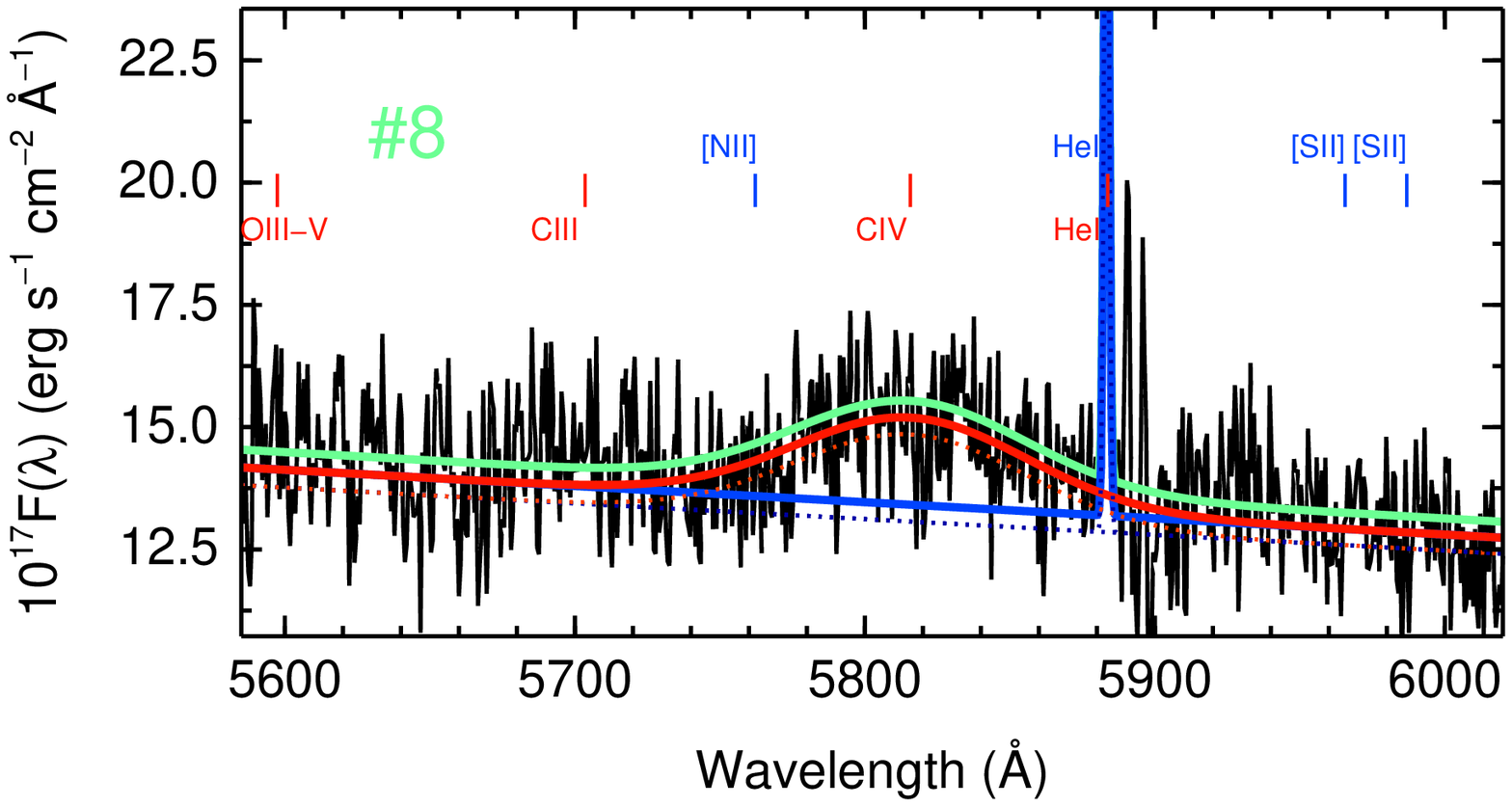}\\

\textbf{Fig. \ref{spectrawr}.} continued. 
 \end{figure*}

%
%

\subsection{Composition of the Wolf-Rayet star population \label{seccompowr}}

\subsubsection{Spectral fitting}

Estimating the number of Wolf-Rayet stars in a given extragalactic spectrum is not straight-forward. One typically depends on only one spectral feature (most likely the \emph{blue bump}) or two at most to trace an assorted zoo of stars in terms of the existence, strength, and width of features in their spectra. The most popular strategy has probably been comparing the observed spectra with the predictions of models of stellar populations like those presented by \citet{Schaerer98}.

The profiles of the extracted spectra for NGC\,625 at the \emph{blue} and \emph{red bumps} are
presented in Fig. \ref{spectrawr}. They are complex, showing a diversity of stellar features (pointing towards
different W-R populations for the different apertures) on top of several nebular lines.
Additionally, some sky background residuals for the  \ion{Na}{id} are also seen at 5893~\AA.
Although still limited to the use of only these two bumps, the quality (i.e. signal-to-noise and spectral resolution) of the spectra is good enough to attempt the more demanding (but more informative) approach of estimating the number of stars as a combination of templates for individual stars. In particular, since the metallicity of \object{NGC\,625} (see Tab. \ref{datosbasicos}) is $\sim$30\% higher
than that of the SMC and $\sim$40\% lower than that of the LMC \citep[i.e. $12+\log(O/H) = 8.03$ and
$12+\log(O/H) = 8.35$, respectively][]{rus92}, and the flux of the Wolf-Rayet features decreases with metallicity, we based our estimation on the results provided by \citet{cro06} for LMC and SMC Wolf-Rayet stars.
One can, with our spectra, distinguish some of the different types of Wolf-Rayet. However, they are not (yet) comprehensive enough to disentangle all the different templates provided by \citet{cro06}. Because of that, we ran different ensembles of fits, each based  on only a subgroup of three templates: an early WN (i.e. WN2-4), a late(-ish) WN, which could be  WN5-6 or WN7-9, and a "WC" (encompassing WC4 and WO). The different subsets of templates used are listed in Table \ref{setoffits}. \citet{cro06} provide in their tables 1 and 2 mean optical luminosities for the different stellar features including or excluding binaries. The fraction of binaries among the WR for the SMC and the LMC is 40\% and 30\%, respectively \citep{Foellmi03a,Foellmi03b} and it is reasonable to expect a similar fraction for \object{NGC~625}. Thus, we used in our templates the mean fluxes when including binaries. The effects of using the mean values when including only single stars, will be discussed below.

\begin{table}
\caption{Subsets of templates by \citet{cro06} used in the fits.}       
\label{setoffits}      
\centering          
\begin{tabular}{ccccc}     
\hline\hline
Subset &     
Metallicity  &
early WN    &
late  WN    &
WC \\
\hline 
A & LMC & WN2-4 & WN5-6 & WC4$_{\rm bin}$\\
B &          & WN2-4 & WN7-9 & WC4$_{\rm bin}$\\
C &          & WN2-4 & WN5-6 & WO$_{\rm bin}$\\
\hline
D & SMC & WN2-4$_{\rm bin}$ & WN5-6 &WO$_{\rm bin}$\\
\hline                
\end{tabular}
\end{table}
%
%

\begin{table*}
\caption{Results from spectra modeling and HST photometry.}       
\label{numberwr}      
\centering 
\small       
\begin{tabular}{|c|c|cc|cc|cc|cc|c|c}     
\hline\hline
 &
 &     
\multicolumn{2}{c|}{Subset A (LMC)} &
\multicolumn{2}{c|}{Subset B (LMC)}  & 
\multicolumn{2}{c|}{Subset C (LMC)}  & 
\multicolumn{2}{c|}{Subset D (SMC)}  &\\
\hline
 Aper. &
 FWHM &
 E/L/C & WR$_{tot}$ & 
 E/L/C & WR$_{tot}$ & 
 E/L/C & WR$_{tot}$ & 
 E/L/C & WR$_{tot}$ & 
 n$_{HST}$\\
  &
 (\AA) &
  & & 
  & & 
  &  & 
  &  & 
 \\
\hline                    
$\sharp$1 & 28 & 4 / 0 / 1 &  5 & 4 / 1 / 1 & 6 & 3 / 1 / 5 & 9 & 23 / 0 / 3 & 26 &  \ldots\\
$\sharp$2 & 16 & 0 / 1 / 0 &  1 & 1 / 1 / 0 & 2 & 0 / 1 / 0 & 1 & 0 / 3 / 0 & 3 & 3\\
$\sharp$3 & 23 & 1 / 5 / 0 &  6 & 6 / 6 / 0 & 12 & 1 / 5 / 1 & 7 & 0 / 22 / 1 & 23  & \ldots\\
$\sharp$4 & 33 & 8 / 0 / 2 &  10 & 7 / 1 / 2 & 10 & 9 / 0 / 15 & 24 & 33 / 0 / 10 & 43 & 11\\
$\sharp$5 & 45 & 2 / 0 / 1 &  3 & 3 / 2 / 0 & 5 & 3 / 0 / 5 & 8 & 16 / 0 / 2 & 18 & 2\\
$\sharp$6 & \,7 & 1 / 0 / 0 &  1 & 0 / 1 / 0 & 1 & 1 / 0 / 0 & 1 & 0 / 2 / 0 & 2 & 5\\
$\sharp$7 & 26 &  1 / 0 / 0 &  1 &  1 / 0 / 0 &  1 &  1 / 0 / 0 &  1 &  0 / 2 / 0 &  2 & 6\\
$\sharp$8 & 70 & 0 / 0 / 1 &  1 & 0 / 0 / 1 & 1 & 0 / 0 / 4 & 4 &  0 / 0 / 4 & 4 & 4\\
\hline
WR Total   & &17 / 6 / 5 & 28 & 22 / 12 / 4 & 38 & 18 / 7 / 30 & 55 & 72 / 29 / 20 &  121    & \ldots\\
\hline
O Stars &  &\multicolumn{6}{c|}{2850 }    & \multicolumn{2}{c|}{1995 }  & \\
\hline
WC / WN &  &\multicolumn{2}{c|}{0.22}  & \multicolumn{2}{c|}{0.12}  & \multicolumn{2}{c|}{1.20}  & \multicolumn{2}{c|}{0.20}  & \\
10$^3$ WR / WO &  &\multicolumn{2}{c|}{10}  & \multicolumn{2}{c|}{13}  & \multicolumn{2}{c|}{19}  & \multicolumn{2}{c|}{61}  & \\
\hline                
\end{tabular}
\tablefoot{
Column 2: Assumed width for the stellar features; Columns 3-6: Estimated number of W-R and O stars (E=early WN; L=late WN; C=WC); Column 7: Number of massive stars in the HST images.
}
\end{table*}

Fits of the \emph{red} and \emph{blue bumps} were done separately. The ingredients and procedure for the fits using a given set of templates were as follows:

\begin{itemize}
\item In addition to the stellar features, we included in the fit a 1-degree polynomial representing the emission of the underlying stellar population and a set of nebular lines (\feiii$\lambda$4658, \heii$\lambda$4686, \feiii$\lambda$4701, \ariv+\hei$\lambda$4712, and \ariv$\lambda$4740 in blue and \nii$\lambda$5755 and \hei$\lambda$5876 in the red). They are marked with blue ticks in Fig. \ref{spectrawr}. 
The \mgi$\lambda$4562.6 and \mgi$\lambda$4571.0, clearly visible in apertures $\sharp$1-$\sharp$3, were not included in the fit. Since they are too blue with respect to the \emph{blue bump}, they did not have any major influence in our fitting technique.
Nebular lines within each spectral range were linked in wavelength and with an imposed width of  $\sigma$=0.7~\AA\, for all of them. Flux was left as free parameter. We obtained an initial guess for the nebular line fluxes using \texttt{splot} within IRAF.

\item Stellar features were modeled by Gaussians with all the features in a bump having the same width. Given the complexity of the fits in the \emph{blue bump}, this width was not a free parameter. Instead, we imposed a different value for each aperture aiming at reproducing the \heii\, line profile as best as possible. The assumed value for each aperture is shown in the second column of Table \ref{numberwr}.

\item Firstly, we modeled  the \emph{red bump} with a  single Gaussian centered at 5808~\AA\footnote{All the wavelengths mentioned in this section are rest-frame.}. Its width was left free, and the flux could be an integer number of times (including 0) the fluxes provided in Table 2 of \citet{cro06}, scaled at the distance of \object{NGC\,625} and assuming no intrinsic reddening for the galaxy (see Paper II, in prep).  
We obtained a first estimation of the number of WC (or WO) stars from the model that minimised the $\chi^2$. 
\item Secondly, we modeled the \emph{blue bump}, using Gaussians representing the features of \nv$\lambda$4610, \niii$\lambda$4640, \ciii$\lambda$4647-4650/\civ$\lambda$4658  and \heii$\lambda$4686. For the WC(WO) stars, \citet{cro06} give the global integrated flux of the last three features. After trying different relative ratios between the carbon and helium features, we adopted 0.8:0.2 as the most adequate one to reproduce the observed profiles. Then, we reproduced the flux of all these features as a combination of an integer number of WN and WC (or WO) stars, again allowing for non-detection of WN stars. While the number of WN stars was left free, for the number of  WC(WO) we used the results for the \emph{red bump}, allowing only for small
variations in the number of stars. 
The model that minimised the $\chi^2$ gives an estimation of the number of stars for each type: early-type WN, late-type WN and WC.

\item Thirdly, we compared the results from the \emph{blue} and \emph{red bump}, regarding the number of  WC(WO) stars. Typically they were consistent (i.e. only in a few cases they gave slightly different number of stars). Values presented in Table \ref{numberwr} refers to the number of stars derived from the fits in the \emph{blue bump}. When they differed, we checked that the predictions of the \emph{blue bump} could also reasonably reproduce the \emph{red bump}.

\end{itemize}

\subsubsection{Caveats related to template selection}

Since the metallicity of \object{NGC~625} is between those of the SMC and LMC, we chose the most reasonable set of templates at each metallicity as the basis for the subsequent considerations. These were WN2-4$_{\rm bin}$/WN5-6/WO$_{\rm bin}$ for the SMC (subset D) and WN2-4/WN5-6/WC4$_{\rm bin}$ for the LMC (subset A), where \emph{bin} refers to templates with binary stars taken into consideration. The minimum $\chi^2$ was comparable when using any of the four subsets. Likewise, the spectral profiles were reproduced with similar accuracy.
In the following, we discuss the subtleties associated to the use of other ensembles of templates and why we consider predictions using subsets A and D as the most reasonable.

\emph{Including or excluding binaries:} Studies on Galactic massive O stars indicate that an important fraction of these stars are in a state of binary interaction \citep[e.g. ][]{Sana11,Kiminki12} and similar results have been reported for the Magellanic Clouds \citep{Foellmi03a,Foellmi03b}. Therefore, our preferred subsets of templates were those for which observations of binaries stars were  taken into account. However, when possible we also fitted the spectra using the templates made from observations of single stars only (ensembles not included in Table \ref{setoffits}).
Luminosities for the LMC templates are only slightly smaller and the number of stars estimated was the same independently of whether templates with or without binaries were used.
When using SMC templates based on only single stars, the fits  predict a number of WC $\sim1-2$ larger, depending  on the aperture, but an unrealistically large number of WN stars. 

\emph{WC4 or WO as template for WC:}
 As shown by \citet{Crowther98} when moving from lower to higher excitation (i.e. $\sim$WC4 to WO1), the ratio \ov$\lambda$5590/\civ$\lambda$5808 should increase, being $\lsim$0.01 for WC4 but $\lsim$0.2 for WO4 (and even higher for earlier types). Whilst \civ$\lambda$5808 is clearly detected, there is no hint of \ov$\lambda$5590 in our spectra. Therefore, the presence of WC4 stars over WO stars in NGC~625 is more reasonable. For models at the metallicity of the SMC, there is no WC4 template available, so we had to use the templates for WO stars.  For completeness, we fitted the spectra using the ensemble C that includes the WO template for the LMC, even if we do not expect WO stars to be responsible of the observed spectra. Since fluxes for WO stars are smaller than for WC4 stars, this set of fits estimate typically up to three more times WR stars - depending upon aperture -  than ensemble A. 


\emph{WN5-6 or WN7-9  as template for late WN:}
This dilemma concerns uniquely the fitting using the LMC templates since no WN7-9 template for the SMC metallicity exists. Both types of stars present  \nv$\lambda$4610 and \heii$\lambda$4686, with the \heii\, line being $\sim$6.7 (4.2) times stronger than the \nv\, line for the WN5-6 (WN7-9) stars. 
In view of the spectra in Fig. \ref{spectrawr}, we do not expect to have WN7-9 in apertures $\sharp$2 and  $\sharp$3. Likewise, spectrum in aperture $\sharp$ 6 is not affected by this discussion, since it can be properly reproduced  with only WC stars.
Regarding apertures $\sharp$1, $\sharp$4 and $\sharp$5, the \emph{blue bump} is dominated by early-type WN stars, and considering the WN7-9 or the WN5-6 template as representative of late-type W-N stars changes only marginally the total number of estimated W-R stars.
Finally, due to their lower quality, the fit of the spectra in apertures $\sharp$6 and $\sharp$7 is not reliable and very dependent on the used set of templates. However, for aperture $\sharp$6, emission features are relatively narrow, supporting the existence of late-type W-N stars in this aperture.  For aperture $\sharp$7, even if the spectrum is consistent with the existence of W-N stars, it is not possible to discern a more refined classification with these data.

\subsubsection{Adopted W-R numbers}

The final number of stars per aperture is shown in Table \ref{numberwr} whilst the optimal fit to the spectrum profile when using the templates for the ensemble A is displayed in Fig. \ref{spectrawr} with a green line.

Interestingly, the assumed width for the features in the fits increases from aperture $\sharp$6, continuing for  all the other apertures up to  aperture $\sharp$8 tracing a sequence from apertures where the number of stars is dominated by late-type WN to those dominated by WC. This is very much in accord with the trend observed for individual stars, that typically present FWHM of $\sim$15~\AA\, for WN7-9, $\sim$22~\AA\, for WN5-6, $\sim$30-60~\AA\, for WN3-4 and up to $\sim$70~\AA\, for WC4 \citep{cro06} and supports our method of disentangling the different types of W-R stars.

Fig. \ref{spectrawr} also shows that there is a clear detection of nebular \heii\ in two of the apertures, $\sharp$2 and $\sharp$3, both dominated by late-type WN stars.  Additionally, there is marginal  detection of nebular \heii\ in aperture $\sharp$1. This is relevant since only certain classes of
W-R stars, and maybe certain O stars, are predicted to be hot
enough to ionise nebular \heii\footnote{ionisation potential:  $h\nu = 54.4$ eV.} \citep{Kudritzki02,Crowther07}.
The analysis of this emission will be part of a forthcoming communication devoted to the analysis of the ionised gas (Paper II, in prep). 

Finally, it is worth noting that the spectral profiles for apertures $\sharp$1 and $\sharp$4 are not  well reproduced. This might point towards an underestimation of the \nv\, flux in the templates. Alternatively, it might call for a more refined modeling of the spectral profile than the one devised here.
We envision two possible ways of improving the methodology presented here.
On the one hand,  the extraction of the spectra associated to W-R emission can be improved using more accurate methods than the  use of square apertures as presented in Sect. \ref{subsecpinpointing}. One possibility particularly promising for IFS data is extending the well-established concept of crowded field photometry in images into the domain of 3-dimensional spectroscopic datacubes \citep{Kamann13}. This has been proven to be extremely successful for disentangling the spectra of heavily blended stars \citep{Husser16}.
On the other hand, it is known that luminosities of the spectral features of a given W-R type spread over $\gsim$2 orders of magnitudes. In that sense, fits could be improved by passing from the use mean templates per W-R type to  a more refined grid of models \citep[e.g. PoWR - The Potsdam Wolf-Rayet Models\footnote{\texttt{\href{http://www.astro.physik.uni-potsdam.de/~wrh/PoWR/powrgrid1.php}{http://www.astro.physik.uni-potsdam.de/$\sim$wrh/PoWR/\\powrgrid1.php}}},][]{Hamann04,Todt15,Sander12}.
MUSE, already operating at the VLT is able to provide higher quality data as those presented here in terms of depth and spatial resolution. Likewise, MEGARA at the GTC will collect IFS data at even higher spectral resolution. These and similar instruments will allow to test these two methodologies.


In the following, we discuss our estimations of W-R star numbers with those found in galaxies at similar distances, model predictions and with already published HST photometry for this galaxy. The discussion will be based on the results obtained when using the ensembles A and D.



\section{Discussion \label{secdiscu}}

\subsection{The numbers of W-R stars}

\subsubsection{Comparison with published HST photometry}

The way of estimating the number of W-R stars presented in the previous section represents a step forward with respect to the use of the integrated luminosity of one or two W-R features in the sense that it is able to estimate not only the number of W-R stars, but also the range of type. With the assumed luminosities for the templates,   we estimate the existence of more than one star in most of the apertures, that were unresolved at the VIMOS spatial resolution. Since W-R spectral features for a given type can vary $\gsim$1-2 orders of magnitude in flux, from object to object \citep[e.g.][]{cro06}, and the estimated numbers of stars depends largely of the used set of templates, it would be desirable to test with independent information whether the numbers of stars estimated in Sect. \ref{seccompowr} are reasonable. 
In this way, we can evaluate whether LMC or SMC templates are more adequate to reproduce the W-R emission in \object{NGC\,625}.
An excellent option for doing so, would be the use of the already published high spatial resolution photometry with the HST for this object \citep{can03,mcq12}\footnote{Programme 8708, P.I.: Skillman}.
In this section, we correlate the location of our regions with W-R emission with the positions, $V-I$ colors and $I$ magnitudes of the individual stars kindly provided by Dr. J. M. Cannon.

\begin{figure}[th]
\centering
\includegraphics[angle=0,width=0.48\textwidth,
clip=]{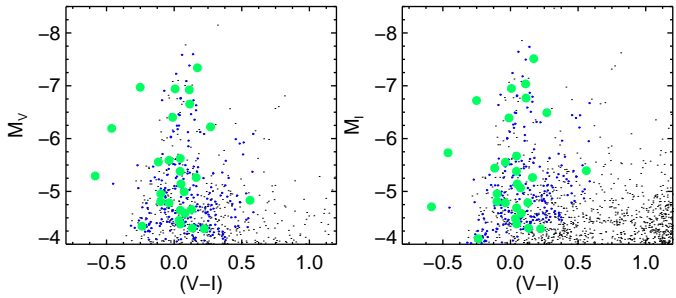}\\
   \caption[HR diagrams with candidates to W-R stars]{A section of the HR diagrams for \object{NGC~625} presented by \citet{mcq12}. Stars in the central part of the galaxy satisfying the criteria of $M_V<-4.05$ and $-1.0<V-I<0.6$ are shown with blue dots. Those stars within the VIMOS apertures are shown with larger green circles. 
 \label{hrdiag}}
 \end{figure}

\begin{figure}[th]
   \centering
\includegraphics[angle=0,width=0.45\textwidth, bb=30 25 750 350,
clip=]{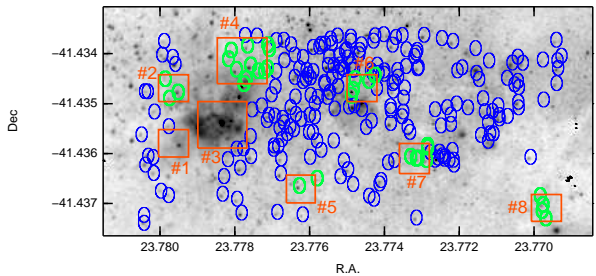}
   \caption[Position of the candidates in the HST image]{
A section of the HST image of NGC\,625 in F555W (programme 8708, P.I.: Skillman).
All the stars in the central region of the galaxy that satisfy the criteria of $M_V<-4.05$ and $-1.0<V-I<0.6$ are marked with blue circles. Those falling within the apertures defined in Sec. \ref{subsecpinpointing} (red squares) are shown in thicker green circles.
There is no available photometry for apertures $\sharp$1 and $\sharp$3.
\label{lochst}}
\end{figure}

The aim is to quantify whether the number of W-R stars that we have estimated in Sect. \ref{seccompowr} is compatible with the number of massive stars found at each location.
This exercise needs accurate astrometry both in the HST images and the VIMOS maps.
We computed the best astrometric solution for the HST archive images by selecting $\sim$40 stars with $M_I\le-6.9$ spread over the whole field of the WFPC2. We used \texttt{ccmap} within IRAF and allowed for shift and scaling in both directions as well as rotation. A few additional,  slightly fainter, stars  in the WF4 camera were also added.

Astrometry for the VIMOS-IFU maps was extremely difficult due to the relatively small field of view and the scarcity of point sources. For pointing 1, we correlated the five peaks of emission in the blue continuum maps with bright stars in a gaussian-smoothed version of $F555W$ HST image. Then we computed the astrometric solution assuming no rotation. For pointing 2, identifying enough point sources was not possible. Therefore, we utilized the solution found for pointing 1 and applied the offset commanded at the telescope.  

Once the positions in the HST images and VIMOS-IFU maps are matched, one can count the number of massive stars in the selected apertures. 
Published $M_V$ magnitudes for W-R stars in SMC are always brighter than $M_V=-4.05$ \citep{bar01,Foellmi03a}. Therefore, we set this value as the lower limit for the $M_V$ magnitude of a putative W-R in \object{NGC~625}.
Likewise, since W-R stars are blue, we restricted our search to stars with $-1.0<(V-I)<-0.6$ colors.
The position in the color magnitude diagram for the stars in the central part of \object{NGC~625} satisfying these two criteria  is displayed in Fig. \ref{hrdiag}  while the locations of the stars in the HST image are shown in Figs. \ref{lochst}, and the number of stars in each aperture are presented in the last column of Table \ref{numberwr}.
We do not include the numbers of massive stars in apertures $\sharp$1 and $\sharp$3 since there is no photometry available for them, most likely due to heavy crowding.

Regarding the other apertures, one would expect a number of massive stars, as found by means of HST photometry, larger or equal to the number of Wolf-Rayet stars, as estimated by means of the VIMOS spectroscopy.
Taking the results using the set of templates A and D as baselines, the number of stars when using the LMC templates is consistent with the HST photometry. We estimate one additional star for aperture $\sharp$5 but this is well within the uncertainties of the method. On the contrary, by using the SMC templates we estimate the same or much larger number of W-R stars, and therefore incompatible with existing HST photometry. 
This is an interesting point. The metallicity of \object{NGC\,625} is closer to that of the SMC, and still modeling using the LMC templates gives more reasonable predictions in terms of number of W-R stars. This might point towards  a non-linear relation between the metallicity  and luminosity of the W-R features. Instead, it would  present an elbow, in the sense that luminosity declines faster at metallicities closer to that of the SMC. Further observations of a larger sample of W-R stars covering a diversity of metallicities would be necessary to test this possibility.

\subsubsection{Comparison with nearby galaxies and model predictions}

Likewise, it would be of interest to see how our predictions, using both LMC and SMC templates, compare with the W-R content in other galaxies at similar distances as \object{NGC\,625} as well as with the predictions of models. For that, in addition to the number of WN and WC stars, we need an estimation of the number of O stars.
This was obtained by co-adding all the spectra in the VIMOS pointings with f(\ha)$>10^{-17}$~erg~s$^{-1}$~cm$^{-2}$ and comparing the integrated \ha\, flux and equivalent width with the predictions of Starburst99 \citep{Leitherer99}, assuming a distance for \object{NGC\,625} of 3.9~Mpc (see Tab. \ref{datosbasicos}) and no extinction intrinsic to the galaxy (Monreal-Ibero et al.  in prep). To be specific, the integrated spectrum was made of the sum of 2481 spectra, that at the distance of \object{NGC\,625} corresponds to an area of 0.398 kpc$^2$, and had an integrated \ha\, flux of f(\ha)$_{int} \sim4.26 \times10^{-12}$~erg~s$^{-1}$~cm$^{-2}$ and an equivalent width, EW(\ha) = 150~\AA.
We used the original Starburst99 models with metallicities $Z=0.008$ (LMC)  and $Z=0.004$ (SMC) and assuming a Salpeter initial mass function (i.e. $\alpha = 2.35$) with M$_{up} = 100$~M$_\odot$.
We also assumed an instantaneous burst of star formation.
The estimated number of O stars as well as the WC/WN and WR/O ratios are presented in Table \ref{numberwr}.

\begin{figure}[!th]
   \centering
\includegraphics[angle=0,width=0.45\textwidth,
clip=]{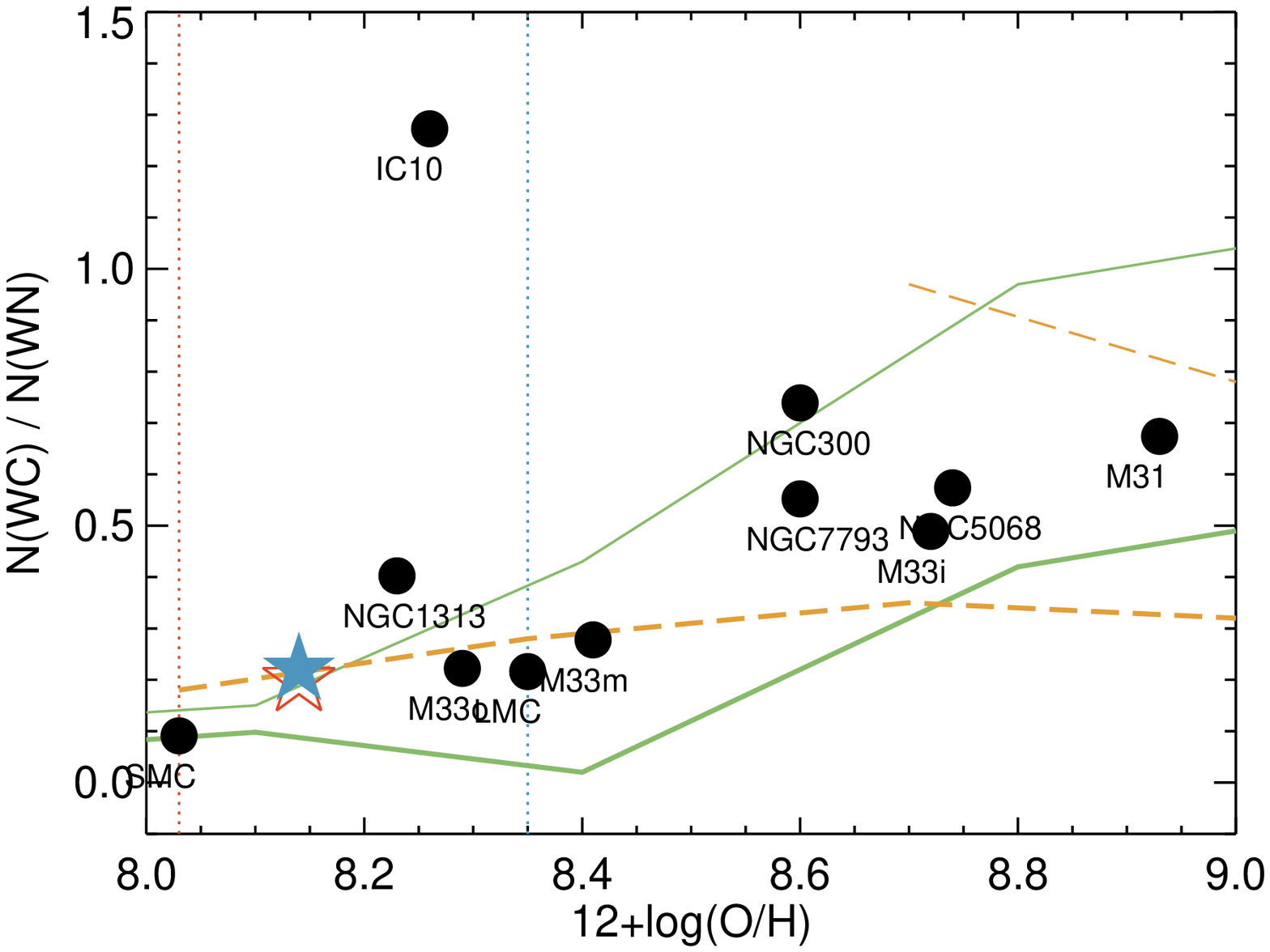}\\
\includegraphics[angle=0,width=0.45\textwidth,
clip=]{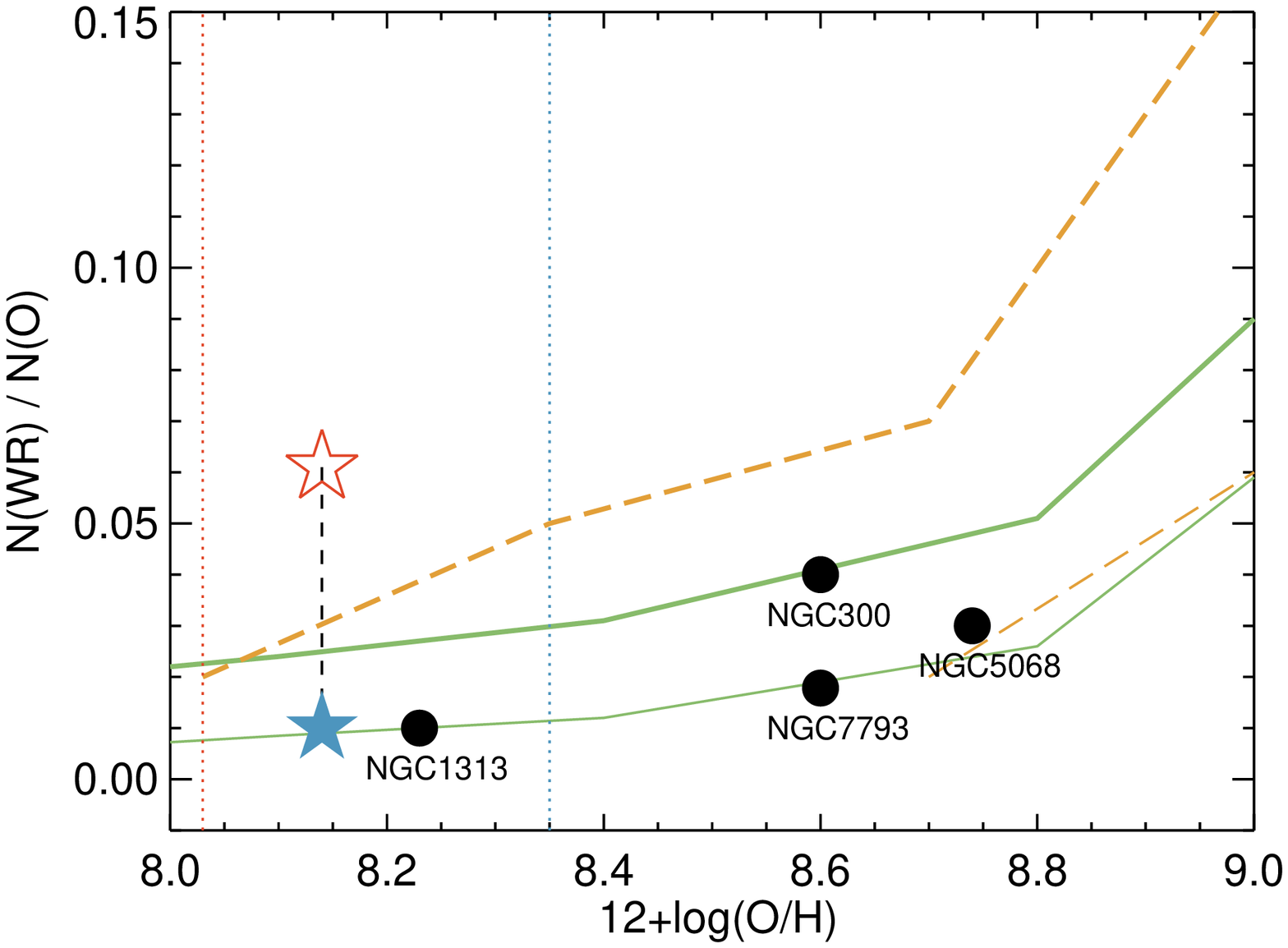}\\

\caption[Relative number of Wolf-Rayet stars versus metallicity]{Relative number of WC to WN stars (\emph{top}) and  WR to O stars (\emph{bottom}) for \object{NGC~625}. Predictions when using the templates for the LMC (subset A) are marked with a filled blue star while whose using the SMC templates (subset D) are marked empty red star. Two vertical lines of these same colors mark the metallicity of the templates.
 We included for comparison the prediction of the models by \citet{Eldridge08} for binary and single stars (thick and thin green lines, respectively) as well as models by \citet{mey05} for stars with rotation ($v$=300~km~s$^{-1}$) and without rotation (thick and thin orange lines, respectively).
Likewise W-R star content for other very nearby galaxies has been added for reference (NGC\,7793, \citet{bib10}; NGC\, 5068, \citet{bib12};  M~31, \citet{Neugent12}; M\,33, \citet{Neugent11}; NGC\,1313, \citet{Hadfield07}; LMC, \citet{Massey15}; SMC \citet{Massey14}, IC~10, \citet{Crowther03}, NGC~300, \citet{Crowther07}).
 \label{wrvsmetal}}
 \end{figure}

Those ratios when using the set of templates A and D are displayed in Fig. \ref{wrvsmetal} together with the estimated W-R content for galaxies at similar distances. Likewise, we superimposed the predictions from the models by \citet{mey05} and the \emph{Binary Population and Spectral Synthesis}\footnote{\texttt{\href{http://bpass.auckland.ac.nz/}{http://bpass.auckland.ac.nz/}}}  \citep[BPASS, ][]{Eldridge08} to address  the role of rotation and existence of binary stars.
The proportion of WC to WN is comparably independent of the used set of templates and of the order of the proportions found for galaxies at similar metallicities, with the exception of \object{IC~10}, for which an anomalous N(WC)/N(WN) ratio has already been reported \citep{Royer01}.
Also, at the metallicity of \object{NGC~625} the model predictions are fairly similar, and thus they do not allow to favor binary stars vs. rotation (see Fig. \ref{wrvsmetal}). However, independently of the set of used templates, our N(WC)/N(WN) ratio is clearly compatible with model predictions, again supporting the approach presented in Sect. \ref{secresults}.

The lower graphic in Fig. \ref{wrvsmetal} displays our estimated proportions of O vs. W-R stars.  The predictions are different by a factor of $\sim$7 depending on whether we use the templates for the LMC or for the SMC. Our estimation using the LMC templates is much more in agreement with model predictions and with the only galaxy at similar metallicities (i.e. \object{NGC\,1313}), than that using the SMC templates.
This again supports the hypothesis of a non linear relation between the metallicity of the star and the luminosity of the W-R features.

All in all, the most reasonable estimation for the W-R content in \object{NGC~625} is that obtained when using the LMC templates (subset A): we estimate a total of 28 W-R stars in the galaxy, out of which, 17 are early-type WN, six are late-type WN  and five are WC stars.

\subsection{The W-R star population and the spatially resolved star formation history  in \object{NGC~625}}

Given their short duration, Wolf-Rayet stars can pinpoint at a very detailed level the moment and  location where recent star-formation has taken place, and thus trace how this propagates through the galaxy. In principle, if a star passes through a W-R phase, the WN stage precedes the WC stage \citep{Crowther07}. For the more refined classification used here, we can consider a "baseline" evolutionary path O star $\rightarrow$ late-type WN  $\rightarrow$ early-type WN  $\rightarrow$ WC. Details and temporal scales for an individual star would depend on its mass, metallicity, amount of rotation, whether it is a binary or not, etc., with multiple variations of this baseline. For example, some stars might not necessarily become WC stars whilst others might suffer LBV episodes. The relevant point here is that roughly speaking, one would expect WN stars to be present predominantly in the younger regions, whilst WC stars should be expected in somewhat older regions \citep[][]{Westmoquette13}. 

According to the results presented in Table \ref{numberwr}, aperture $\sharp$8 (towards the west, and the furthest from the peak of emission in \ha) contains only one WC star. In contrast, apertures  $\sharp$1, $\sharp$5 and $\sharp$4 (closer to the peak of emission) contain a mix of WC and early WN stars.  Aperture $\sharp$7 contains star(s) of type WN.
Finally apertures $\sharp$2, $\sharp$3, and $\sharp$6 contains only WN, and are dominated by late type WN, with only 1 early-type WN in aperture $\sharp$3, the one centered at the peak of emission in both \ha\, and the stellar continuum, and aperture $\sharp$6 is associated to one of the strongest \hii\, regions in the galaxy \citep[NGC\,625 B, in the nomenclature used by][]{can04a}. Thus, the W-R population would suggest that star formation has propagated from the western and more external parts of the galaxy inwards, with most recent star formation at the peak of emission. 
This is crude map of the propagation of the star formation, but roughly compatible with the findings using high spatial resolution imaging with the WFPC2/HST \citep{can03,mcq12}.

In this context, one would expect no WR feature or only a WC star signature in the other brightest region in the galaxy \citep[NGC\,625 C in the nomenclature used by][]{can04a}. In that sense, the non-detection of WR features at this location is reasonable.
However, to check whether the lack of detection of W-R features in this location  was an effect of our imposed threshold for the detection of the bumps, we  extracted the spectrum associated to this region. It did not show any strong sign of blue or red bump.
This nicely supports the interpretation presented above even if this does not reject the possibility of existence of a few W-R stars at this location with line fluxes lower than the average (which is what it is considered in the templates).

Likewise, an alternative (or complementary) scenario to explain why the W-R star identified in aperture $\sharp$8 have been found that far from the rest of the W-R population is the possibility of being a runaway star. One of the two main proposed mechanisms by which a star can be ejected from its birthplace are the supernova explosions. \object{NGC\,625} actually displays non-thermal synchrotron radiation in one of its main \hii\, regions \citep[NGC\,625 C in the nomenclature used by][]{can04a}, which is at $\sim$250~pc north-east of our aperture $\sharp$8.
Supernova remnants can be detected in the radio for up to $\sim10^5$~yrs \citep{Bozzetto17}.
Thus, independently of the age of the WC star, if we assume that the SN was the agent causing the star to run away, then an extremely large kick velocity would be needed for the star to reach is current location. Therefore, the dynamical
ejection scenario, whereby the ejection is caused by encounters
in a dense cluster appears to be more plausible. In this scenario, and assuming an age of $\sim5$ Myr  for the WC star in this aperture, a kick velocity of $\sim50$~km~$^{-1}$ would have been enough for the star to reach its current location.
Alternatively, the star could come from the largest \hii\, region at $\sim$400~pc, if we take as reference the peak of emission in \ha.
For the distance between the WC star and the \hii\, region and assuming a similar age or somewhat younger ($\sim3-5$~Myr), the kick velocity would be $\sim80-130$~km~s$^{-1}$. These are all within the range of velocities for runaway stars suggested by \citet{Rosslowe15}.

\section{Conclusions \label{secconclu}}

We have carried out a detailed study of the central $\sim1\,100\times550$~pc of
the nearby BCD \object{NGC~625} using VIMOS-IFU
observations.
In this paper, we present the data and
discuss the W-R star population.
Our main conclusions can be summarized as follows:

\begin{enumerate}
\item The Wolf-Rayet star population is spread over the main body of the galaxy and does not necessarily coincide with the overall stellar structure. Likewise, while the \emph{blue bump} emission, characteristic of WN stars is more concentrated, with a maximum at the peak of emission in both continuum and \ha, the \emph{red bump} emission is more  extended and peaks $\sim5\farcs0$ north. Indeed, WC emission has been detected at distances up to 400~pc from the  peak of emission in continuum and \ha.

\item Nebular \heii\ has clearly been detected in two of the apertures, $\sharp$2 and $\sharp$3, which are the two dominated by late-type WN stars.  Additionally, there is marginal  detection of nebular \heii\ in aperture $\sharp$1. 

\item The best estimation for the W-R content in \object{NGC~625} is that obtained when using the LMC templates (subset A): we estimate a total of 28 W-R stars in the galaxy, out of which, 17 (61\%) are early-type WN, six (21\%) are late-type WN  and five (18\%) are WC stars. Most of them (36\%) are located in aperture $\sharp$4, at $\sim5\farcs0$ north from the peak of emission in continuum and \ha.

\item The width of the stellar features is correlated with the type of W-R stars found in each aperture, tracing a sequence from those apertures  dominated by late-type WN stars, that present narrower features, to those dominated by WC stars, with broader features. This is very much in accord with the trend observed for individual stars.
      
\item A comparison of our results with HST high spatial resolution photometry, with the W-R content in galaxies at similar distances and metallicites and with the prediction  of stellar evolutionary models indicates that modeling using LMC templates gives a more reasonable number of W-R stars than the use of SMC templates whilst the metallicity of the galaxy is closer to that of the SMC. This suggests that the relation between metallicity and luminosity of the W-R spectral features is non-linear with a steep decline  at lower luminosities.

\item The distribution of the different types of WR in the galaxy is roughly compatible with way star formation has propagated in the galaxy, according to previous findings using high spatial resolution with the HST.
      
\end{enumerate}
   
The methodology presented here to count and characterise the W-R star population in a galaxy constitutes an step forward with respect to previous approaches, in the sense that it disentangles - at least -  three different basic types of W-R stars: early-type WN, late-type WN and WC (or WO), and in that sense it constitutes the path to follow to make the most of the high-quality IFS data generated by instruments like MUSE. Nonetheless, there is room for improvement. At the level of spectral extraction, crowded field 3D spectroscopy \citep{Kamann13} can identify much better the emission associated to a given W-R star candidate. At the level of spectral modeling, a finer grid of templates would help to better constrain the composition of the W-R population. That would be of particular interest in the analysis of IFS data at higher spectral resolution (e.g. R$\sim$10\,000), as those that will be provided by instruments like MEGARA or WEAVE, since at this resolution the different spectral features composing the \emph{blue bump} will be better decomposed.

\begin{acknowledgements}

We thank John M. Cannon and Kristy McQuinn for making the HST photometry of \object{NGC\,625} available to us.
We also thank the referee for the useful comments that have significantly improved the first submitted version of this paper.
Based on observations carried out at  the European Southern           
Observatory, Paranal (Chile), programme 086.B-0042.
A. M.-I. grateful to ESO Garching, where part of this work was carried out, for their hospitality and funding via their visitor program. 
A. M.-I. acknowledges support from the Spanish PNAYA through project AYA2015-68217-P and from BMBF through the Erasmus-F project (grant number 05 A12BA1).
M. R. acknowledges support by the research projects AYA2014-53506-P from the Spanish Minister de Econom\'{\i}a y Competitividad, from the European Regional Development Funds (FEDER) and the Junta de Andaluc\'{\i}a (Spain) grants FQM108. 

This paper uses the plotting package
\texttt{jmaplot}\footnote{\texttt{
http://jmaiz.iaa.es/software/jmaplot/current/html/\\jmaplot\_overview.html}},
developed by Jes\'us Ma\'{\i}z-Apell\'aniz. This 
research made use of the NASA/IPAC Extragalactic 
Database (NED), which is operated by the Jet Propulsion Laboratory, California
Institute of Technology, under contract with the National Aeronautics and Space
Administration.

\end{acknowledgements}

\bibliography{mybib_aa}{}

\end{document}